\documentclass{article}
\usepackage[utf8]{inputenc}

\usepackage{physics}
\usepackage{geometry} 
\usepackage{graphicx}

\begin{document}

\title{\flushleft \bf Do Robots powered by a Quantum Processor have
the Freedom to swerve?\\

{\large On Consciousness, Feelings, Agency and Quantum Artificial Intelligence}
}
\date{}
%\author{\flushleft \normalsize Hartmut Neven$^{1}$, Peter Read$^{2}$, %Leah Willemin$^{3}$, and Tobias Rees$^{3}$\\
%\\
%$^{1}$Google Quantum AI, Venice, CA 90291, United States \\
%$^{2}$Technology investor, London, W1U1QY, United Kingdom\\
%$^{3}$Berggruen Institute, Los Angeles, CA 90013, United States\\
%}

\maketitle
\thispagestyle{empty}
\vspace{-1.6cm}
{\flushleft {\large Hartmut Neven$^{1}$, Peter Read$^{2}$, and Tobias Rees$^{3}$}
\vspace{.3cm}
\\

$^{1}$Google Quantum AI, Venice, CA 90291, United States \\
$^{2}$Technology Investor, London, W1, United Kingdom\\
$^{3}$Berggruen Institute, Los Angeles, CA 90013, United States\\
\vspace{.8cm}
}
{\hfill “... some things happen of necessity, others by chance, others through our own agency ...”

\vspace{.5cm}
\hfill \parbox{11cm}{Epicurus, 341-271 BC, proposed the idea of “atomic swerve", which \mbox{suggests} that atoms may deviate from their expected course, thereby \mbox{permitting} \mbox{humans} to possess free will in an otherwise deterministic universe \cite{Epicurus}.}
\\
\\
\\
\noindent Any scientific attempt to explain consciousness is tasked with reconciling the third person objective perspective of science with our first person subjective experience of the world. 
A good point of departure is to consider situations in which these two perspectives are correlated. We note that behaviors conducive to our well-being, i.e. conducive to maintaining homeostasis, tend to be associated with feelings of pleasure while actions that threaten our homeostasis tend to coincide with unpleasant feelings. We choose a materialist/physicalist approach and find the simplest explanation for this correlation arises from the assumption that we possess agency. If a system has the agency to choose an outcome then presumably it would choose a pleasant over an unpleasant one. Agency implementing preferences manifests itself as a force in a third person theory. If a preference is strong it gives rise to a force causing a deterministic evolution, while weak preferences give rise to non-deterministic evolutions. Quantum physics assigns a high degree of non-determinism to large systems of qubits, a Knightian uncertainty, in which one can not even assign probabilities to observed outcomes. This can give engineered systems the freedom to act on their preferences in ways not always predictable by an outside observer. Our considerations lead us to propose a three part design for an engineered animat for which one may provocatively argue that it is conscious and possesses agency and feelings. i) Drawing on insights gained from a recent experiment demonstrating beyond classical computational capabilities we sketch a quantum circuit that prepares a state for which we argue it may be experienced as pleasurable by a system that assumes it. ii) Employing techniques from quantum error correction we suggest a homeostatic loop allowing the system to protect the pleasant state against perturbations. We also endow the system with the ability “to report feelings''. We do so by constructing a mapping between the space of states seen by the interoception sensors of the animat which measure properties of the quantum state it seeks to stabilize and a space obtained from a word embedding of English language words describing feelings. iii) We integrate this homeostatic engine with a neural network such that incorrect classifications result in forces that threaten the stability of the pleasant state. We conjecture that a system concerned with maintaining homeostasis while learning is a better learner as compared to a system without a homeostatic engine. 
}

\section*{Introduction}
\label{section1}
Neither physics nor computer science nor neurobiology has a good model for conscious experience. We define consciousness as the phenomenon of experiencing the world from a first person subjective perspective. It is challenging to move our understanding of consciousness from philosophical speculation to the realm of experimental science. This is because by definition, in order to describe the conscious experience of a system other than ourselves we have to access its inner experience. This seems impossible and we may give up right here. Of course if we do so then we concede science is mute on what is arguably a highly relevant phenomenon. For instance, more people are interested in the question of what they may experience when they die than in the question how to integrate general relativity with quantum mechanics. This essay suggests avenues to reconcile our experience of the world from a first person perspective with the worldview described by the third person perspective of science. 

A good starting point is to consider situations in which the first person perspective and third person perspective are correlated. We note the following is the case. 
It  appears  that  voluntary behaviors  that  are  good  for  sustaining  our  organism  and  the  proliferation  of  our  species  are  associated with  a  pleasant  sensation.   For  example,  eating  a  sweet  fruit,  taking  a  warm  shower or having a  good  sleep  in cuddly blankets are all activities that feel good.  Inversely, if I engage in activities damaging to my body such as getting too close to a fire then this is associated with an unpleasant, painful feeling \cite{JamesWilliam}.\footnote{The  correlation  between  behaviors  that  feel  rewarding  and  those  sustaining  our  well-being  is  not  perfect. For instance, a person engaged in sport may incur muscle strain but doing regular sport is healthy for the body. A similar example is that many students find it stressful to study hard. Such deviations can explained by the fact that humans can learn to maximize not just the immediate reward but rather the sum of future discounted rewards. There are other cases in which this association goes astray.  It may have been good for a stone age hunter to eat plenty of sweet fruit when he could get a hold of it and enjoy its taste.  But in today’s environment in which sugar laden foods are abundantly available,  the pleasant feeling encouraging what used to be a sustaining behavior now  can  lead  to  health  problems.   So  if  we  adopt  the  perspective  of  evolutionary  psychology  then  we would argue that this correlation was even stronger when our ancestors were still roaming the savannas but  that  the  recent  rapid  changes  in  our  environment  caused  this  relation  to  get  a  bit  out  of  sync. Evolution  when  closer  to  a  steady  state  seems  to  map  productive  behaviors  more  perfectly  to  pleasant feelings.  This observation is worth noting and also requires explanation. This  observation  also  addresses  another  possible  objection,  namely  that  taking  drugs  such  as  opioids  cause  pleasant sensations  yet  are  not  conducive  to  survival.  During  most  of  human  evolution  substances  that  directly  interfere  with  our body’s reward systems were not readily available.} If an organism faces a choice between two actions, one leading to an outcome associated with a pleasant sensation  while  the  alternative  leads  to  an  unpleasant  sensation,  then  the  organism  is  likely  to  choose the  action  leading  to  the  pleasant  outcome.  Most  of  us  would  report  that  they  are  acting  in  this  way. One of psychology’s basic experiments, the Skinner Box, suggests that other organisms such as lab rats act in this manner as well \cite{KrebsJR}. Why is this so? One may jump to answer: “Obviously this helps with self-preservation. This has to be so.”\cite{JamesWilliam}\cite{Popper1978} But this can not be the complete answer. Evolution would work just fine if this correlation were inverted because selection is based on phenotype and does not have access to the inner experience of an organism. Thus evolution would still select for organisms that engage in behaviors conducive to their survival even if they feel unpleasant to them. There must be a different reason. We find that the simplest explanation results from assuming free will exists i.e., that we have the agency to act in accordance with our preferences. If it were the case that a system had the agency to choose one of several outcomes then presumably it would use this freedom to select the state it experiences as rewarding. 

At first glance the laws of physics do not seem to allow for agency. But this is not so and we will show that physics leaves room for agency to exist even though the laws of physics tend to camouflage this possibility. The following example is helpful to discuss agency. Imagine we meet a child who loves vanilla ice cream. Every time we offer her several flavors to choose from she will invariably pick vanilla. So from a third person perspective we could describe her behavior in deterministic terms. A third person observer would say the child acts like an automaton and one can perfectly predict her actions. Still the child could insist that she acts in accordance with her preferences and that she does not care whether her actions are predictable. This is the position of compatibilism \cite{Coates}, namely that free will can exist even in a deterministic world. Let us change the example a little. We roll a die and the outcome will determine which ice cream flavor the child gets to pick. Now the third person description will have to be formulated in probabilistic terms. But here we would not speak of agency. The child does not get to choose her preferred flavor. So we can see that a theory that assigns probabilities to different observable outcomes, such as quantum mechanics, does not automatically allow for agency to exist. Still, we would like to propose the following idea. A force in physics can be seen as an expression of agency realizing a preference. If the preference is strong and consistent it gives rise to a deterministic description and if the preference is weaker and less consistent the description becomes probabilistic. Most recent developments in physics, such as insights gleaned from experiments at the Google Quantum AI lab, open yet more room for agency to be present. A system of about $100$ quantum bits, qubits, evolving under the operation of discrete gates is so complicated that it becomes impossible to predict its outcomes even probabilistically. This gives rise to Knightian freedom, named after the economist Frank Knight \cite{Frank_Knight}\cite{aaronson2013}. It is characterized by a situation in which we can not even make probabilistic predictions. It is now impossible for us to exclude that agency was at play in selecting an outcome. 

This allows us to answer the question “Could a technology be built in principle that perfectly predicts the actions of a human being?”\cite{deutsch2013constructor} The answer is "no". 100 qubits are barely sufficient to describe the electronic structure of a protein molecule. In fact many more qubits are needed. Predicting the molecules' exact behavior likewise is out of reach for any classical computer that can ever be built. A quantum computer can help with such predictions provided the number of outcomes is sufficiently small. Then we can simulate the system in question, evolve it many times, keep a tally of the outcomes and turn them into probabilistic predictions. But often the number of outcomes is combinatorially large. In general the number is $2^n$ where n is the number of qubits. In this case keeping a tally of the observed outcomes becomes impractical. None of this affirmatively proves that agency does exist but it clearly shows that the laws of physics do not exclude it either. Combining this insight with the first person observation that we tend to choose outcomes that feel rewarding and avoid those that feel unrewarding invites the conclusion that agency is the cause for the correlation we sought to explain. 

What about cases in which we observe phenomena that violate the known laws of physics? A good example is the movement of the arms of spiral galaxies which violate Kepler’s laws. In such cases we might be tempted to say that we witnessed an expression of agency. But of course the program of science will proceed to refine the third person description, for instance by introducing the notion of dark matter until any deviations from the predictions of the new model become statistically insignificant. Thus the descriptions of the world from the first and third person perspectives will become more and more congruent leading to psychophysical parallelism. A physicist may be reminded of the fact that descriptions relative to two different inertial coordinate systems are related by well understood transformations. This tends to camouflage that there has been room for agency all along. 

\section*{Consciousness as a ubiquitous immanent property of matter}
\label{section2}
Any attempt to make progress on the scientific treatment of consciousness ought to begin with taking stock of our repertoire of methods that can probe the inner perspective of systems we observe. Let us consider a Gedankenexperiment in which one system may know the experience of another. Let us assume we have two identical copies of a system, A and A’. If system A is in state S and produces an output O=o(S) which is a function of its state then given sufficient perceptual and inferential abilities system A’ can infer by observing O which state S the system A is in. Moreover, since both systems are assumed to be identical, system A’ can know what A experiences by recalling its own experience when being in state S (see Fig. 1) We believe this case, in which two systems are identical and observe each other, is the only scenario that allows for probing the experience of something external.\footnote{In describing this idealized protocol for observing conscious experience we made a number of assumptions: Most importantly we assumed that the systems A and A’ have the sensory and signal processing apparatus sufficient to recognize each others’ outputs and have the faculties of inference and memory to determine state S and recall its associated subjective experience. Moreover we assumed that two systems that are in the same state experience the same. A lesser caveat is that if the output function o does not associate a unique output O with each state S then system A’ can only know that A is in one of several possible states. On more fundamental grounds one could question to what degree it is even possible to have two identical copies of rather complex systems and whether the systems could ever be in the same state given that their environments may never be exactly identical.} 

Humans employ a protocol similar to the one described here when trying to discern the subjective experience of another person. Imagine a mother watching her child eating ice cream. When she sees a satisfied smile on her child’s face, she is rather certain the child finds the ice cream tasty. True, she can not know in precise detail how the ice cream tastes from the child’s perspective but she probably has a good sense of what is going on. Moreover, she can entice her child to communicate additional interpretable outputs by asking the child how the ice cream tastes or what the taste is similar to. In fact such self reporting by human subjects is the only measurement protocol that is available to consciousness researchers to probe conscious experience \cite{Koch}. Of course we are making additional assumptions when mapping the abstract scenario above to the case of two humans. I can not verify that anything else is similar to me. Taking this realization to the extreme leads to the position of solipsism, namely that I may be the only one with a conscious mind. Solipsism is logically closed and can not be refuted \cite{Solipsism}. Hence all attempts to explain consciousness differently are really efforts to make it plausible that consciousness exists outside of myself.

\begin{figure*}[t]
\centering
\includegraphics[width=\textwidth]{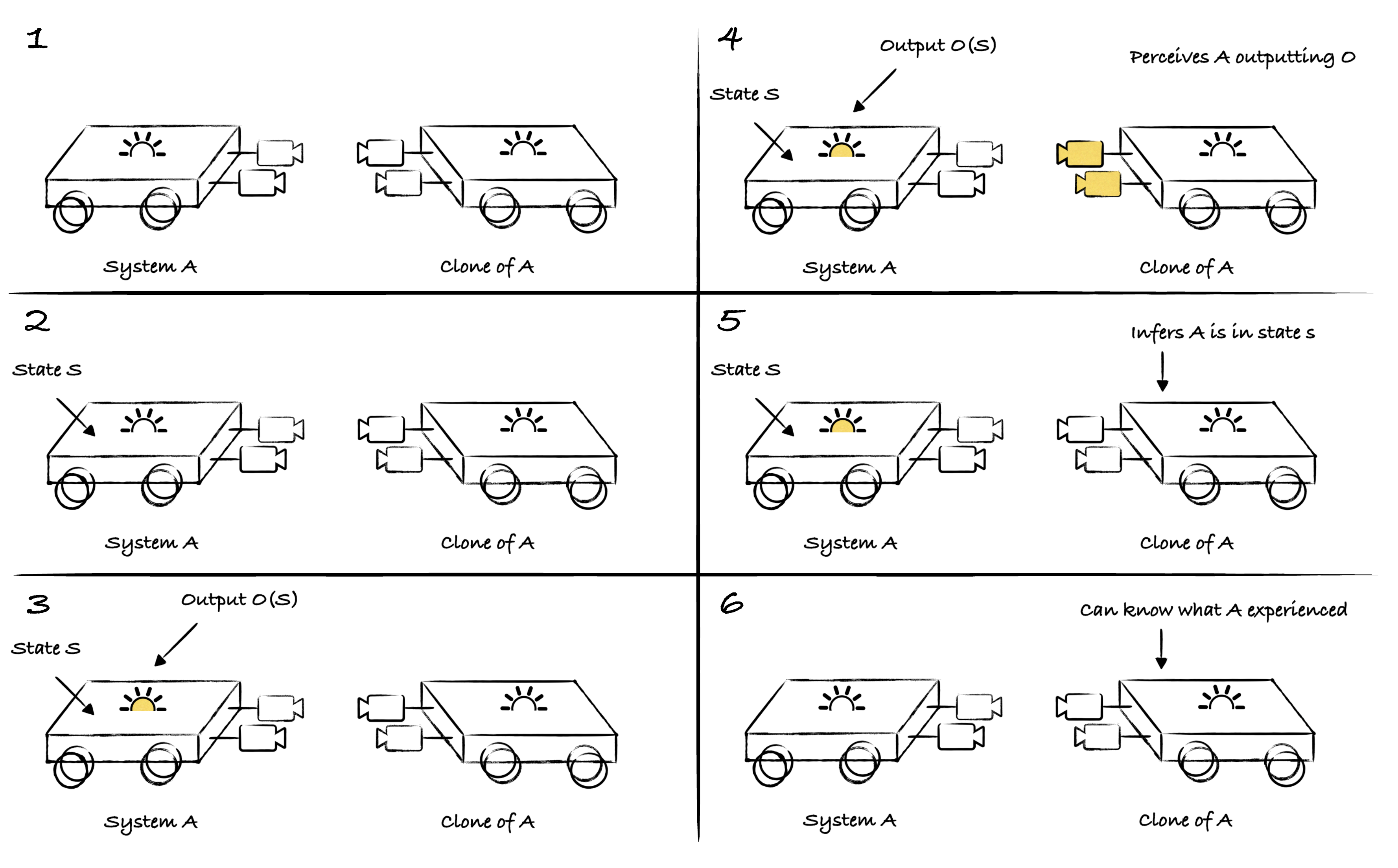}
\caption{Conscious experience is an observable between two systems that are clones of each other. As systems become more dissimilar this protocol breaks down (inspired by \cite{Braitenberg1}).}
\label{fig:1}
\end{figure*}

The more dissimilar two systems are the less they can know about each other’s conscious experience. Consider the following example: assume engineers came up with a new jet engine and based on their calculations predict that it will accelerate the airplane to 1000km/h. It would be straightforward to test this prediction because there is a clear measurement prescription. We would set up two spatial markers separated by a known distance and then measure how much time it takes the airplane to fly from one marker to the other. From this we can infer the speed and test the engineers' prediction. But now let us consider the autopilot. The autopilot reads in various signals coming from the plane’s sensors, and based on those it calculates what thrust the engines or which angles the rudder or flaps should have to ensure a stable flight. We may ask the question: "To what degree is the autopilot conscious?" For example, we could try to answer this question by computing the so-called “$\Phi$-measure” which aims to quantify the amount of consciousness that can be present in a system \cite{Tononi}. Let us assume we obtained a $\Phi$-value of 1000. But now we are stuck. There is no measurement protocol we could use to verify this prediction and therefore it is vacuous in the context of experimental science. Alternative measures to quantify consciousness would suffer from the same predicament.

In short, when it comes to assessing the conscious experience of another creature or system the maxim  “It takes one to know one.” seems to apply. This may be the cause of a widespread misconception which equates the ability to communicate with the propensity for conscious experience. We often read statements like: “I think my fellow humans are conscious. My dog and cat seem conscious as well. I am not so sure about a goldfish but a fungus, tree or a rock definitely do not feel conscious to me.” This ordering correlates well with the ease with which we can relate to the subjective experience of the other being in question.\footnote{A strategy we could think of employing to assess the conscious experience of a system that is dissimilar to another is by creating a sequence of sufficiently similar systems A $\leftrightarrow$ A’ $\leftrightarrow$ A’’ $\leftrightarrow$ … such that any two neighbors can interpret each other’s output. It would be difficult to assess what gets “lost in translation” but it could be a strategy to expand our sense of what another system experiences.} But who is to say that being a rock or an atom does not feel like something? We can not know! 

Even in the face of these limitations to arrive at an experimentally verifiable theory of consciousness we are still able to make non-trivial statements by making an assumption. In particular we want to discuss inferences we can make by subscribing to a materialist worldview which we believe many scientists would accept as rather innocuous. Materialism holds that all phenomena including mental ones are brought about by configurations of matter \cite{Materialism}. In particular, a materialist assumes that there exists a physical correlate of consciousness, a “chunk of matter”, for instance a group of nerve cells or a set of molecules, that is the seat of conscious experience. Comas or vegetative states can be caused by lesions to relatively small regions of the brainstem \cite{Parvizi}. This suggests that conscious experience could be housed in a rather small volume of matter. 

So where does the physical correlate of consciousness reside? Does it exclusively exist in highly evolved biological organs? We prefer the panpsychist notion that consciousness may be a distributed feature of the universe located with the elementals of physics \cite{Panpsychism}. I.e., we should think of consciousness similar to the way we think of spin, mass or electric charge. Logically this is the most parsimonious stance. At this point we can not exclude the possibility that more complex configurations of matter are required to bring about conscious experience, similar perhaps to the intricate mechanisms required to bring about locomotion. We also expect that certain configurations of matter organize elementary conscious moments into more elaborate patterns in the same way that magnetic moments of atoms spontaneously self-organize into a macroscopic ferromagnet. Panpsychism allows for this. 

\begin{figure*}
\centering
\includegraphics[width=0.8\textwidth]{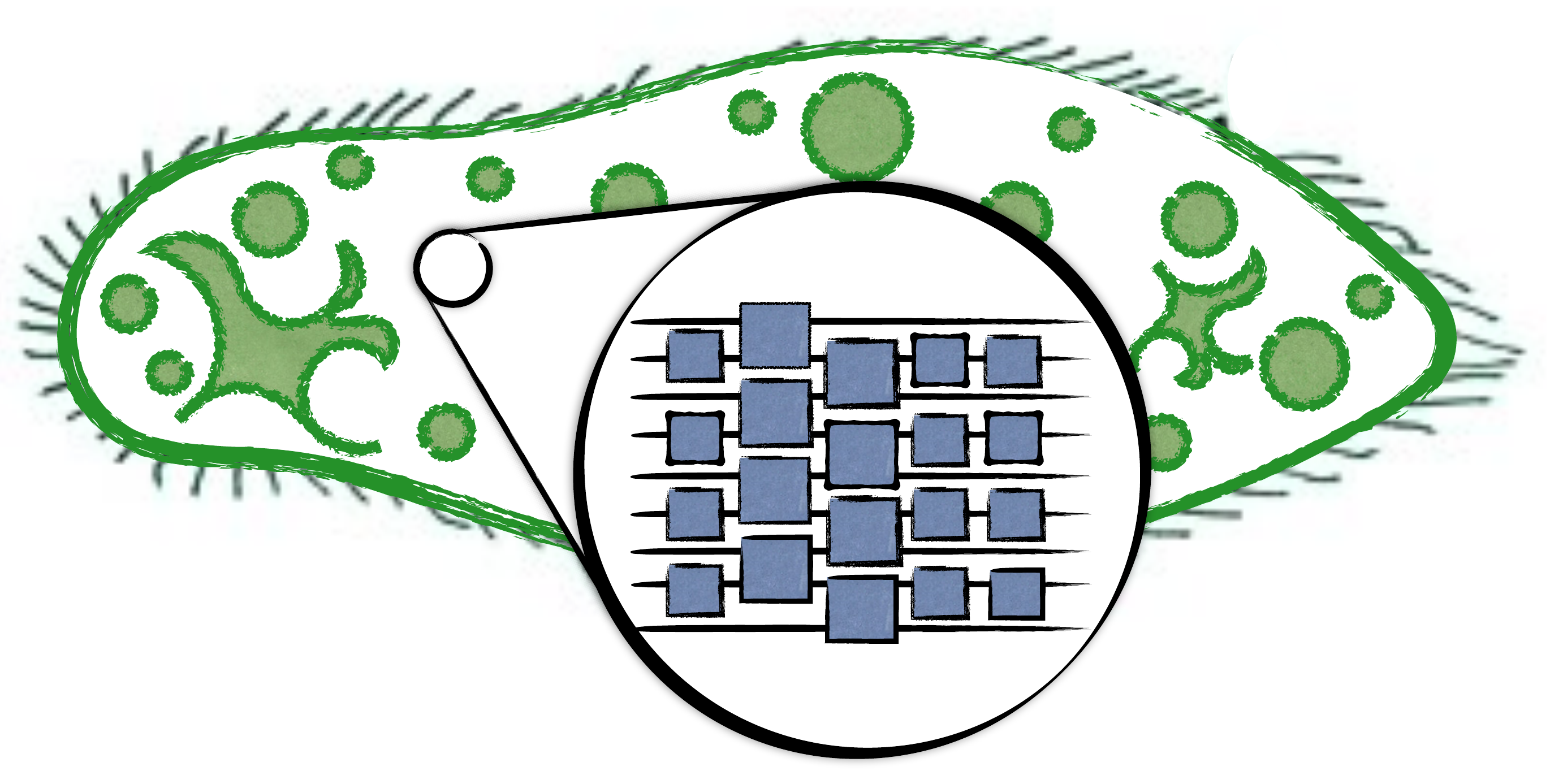}
\caption{Beyond-classical quantum computational resources could be realized in subcellular structures. We may underappreciate the sophistication of the machinery needed to keep a cell alive and to allow it to proliferate. An indication of this is that prokaryotes already appeared about 3.8 billion years ago while mammals appeared 200 million years ago and Homo sapiens just 300 thousand years ago. So far it has also proven elusive to engineer artificial cells able to procreate.}
\label{fig:2}
\end{figure*}
To be definite we want to propose what we would like to call the “Generalized Penrose Hameroff Conjecture”: Consciousness is how a system experiences the emergence of a unique classical reality. To elaborate, conscious experience is associated with a system assuming via decoherence one classical configuration from the multitude of configurations over which the quantum mechanical wave function has support.\footnote{Note that identifying wave function collapse as the physical correlate of consciousness is different than suggesting that consciousness causes the wave function to collapse as von Neumann or Wigner did \cite{vonNeumann}\cite{Wigner}.} The original proposal of Penrose and Hameroff assumes a specific decoherence channel for the quantum mechanical wave function called “objective reduction” \cite{Hameroff}. Objective reduction occurs when the spread of mass associated with the wave function exceeds a threshold. Whether this mechanism does indeed exist remains subject to experimental investigation \cite{Marshall} but we may find that it does not exist. Indeed a recent test of gravity-related wave function collapse ruled out one version of objective reduction \cite{Donadi}.
However, the wave function of any open quantum system is subject to various and in fact much stronger decoherence mechanisms which can play the same role that Penrose assigns to objective reduction. Associating a conscious moment with the collapse of the wave function is appealing because at that moment the system “realizes” or “perceives” which classical reality it inhabits, matching the popular notion that consciousness acts as a receiver or reducing valve \cite{Huxley}. For us then, stated simply, consciousness is what it feels like to select a single classical reality from the multiverse.

One may object to panpsychism on the grounds that one should not use the same linguistic terms when describing the conscious moment an atom may be experiencing when an electron transitions from a higher to a lower energy orbital as the pleasure one experiences when looking at an aesthetic photo of a galaxy taken by the Hubble telescope \cite{Damasio}. For the latter experience to occur a high degree of organization of matter is required. The Hubble telescope shoots a photo, transmits it to earth, it appears on my computer screen, my eyes capture the photons emitted by the screen, our visual cortex processes the signals coming from the retina, neurons in the higher visual cortices encode the percept and broadcast it to other parts of the brain. Somewhere along this pathway a highly orchestrated "wow" experience occurs.

We concur with Koch that consciousness like electric charge may not have a function in sensu stricto \cite{KochFeeling}. But like electric charge can be made useful in a battery, organizing conscious moments in a human brain capable of recursive reasoning may confer functional advantages. Indeed, there is some empirical support for the hypothesis that one function of consciousness may be to produce counterfactual representations of a situation using generative models of the environment and the self \cite{Kanai}. A quantum system, comprised of a superposition of alternative classical configurations, is naturally suited to fulfill this role and it can be exponentially more efficient in executing algorithms needed to perform this task \cite{Lloyd2014}\cite{LaRose}. 

\section*{Agency is permitted by the laws of physics}
\label{section3}
Arguably the most direct way to explain the correlation between pleasant sensations and behaviors conducive to an organism’s well-being is to postulate the existence of free will. If an organism possesses the agency to freely choose a state, presumably a rewarding over an unrewarding one, then the evolutionary process would naturally favor the association of states that feel rewarding with behaviors helping an organism to thrive. Vice versa, in a world without agency, in which actions are determined solely by chance or by deterministic laws, how actions feel is irrelevant to guiding the behavior of an organism. In this case we do not know how to explain how a reward system can work and why evolution associates behaviors conducive to homeostasis with rewarding sensations.
 
From introspection it appears to many humans that indeed we can freely choose which action to take. To most of us sentences such as “I did this because it was fun.” or “I am going to avoid that because it will hurt.” make sense and they can be heard daily. However this sense of agency, the free will to pursue an action leading to a pleasant sensation, may just be an illusion.\footnote{However, we want to counter one argument that is often heard in this context, namely that the equations of motion in physics are deterministic and hence free will can not exist. We would argue that it is the other way around. Science implicitly makes the assumption that some level of non-determinism exists. Otherwise we could not check our theories. A theory can only be falsified by checking its predictions and if we are not free to at least randomly choose experimental settings then we can never be certain that the theory holds universally. This would be like an accountant who would not be allowed to spot check any line of a ledger as she pleases but only certain transactions. So non-determinism may not exist but in that case we can not trust our scientific theories either.}

It is often stated that a necessary condition for free will is that it should be impossible to predict an action of an agent by observing all events in its past light cone, which is physics jargon for the set of all events that can causally influence an action.\footnote{The reason that the impossibility to predict an agent’s action is only a necessary but not a sufficient condition for free will is that if the agent has several choices but they are selected randomly, say by the throw of a die, then we would still not consider these choices to be an expression of free will.} However, the philosophical view of compatibilism \cite{Coates} which defines free will as acting in accordance with one’s motivations, denies that the inability to predict actions is a necessary condition for free will to exist. This position was illustrated in the introduction with the example of the child who always chooses vanilla ice cream. Even though her behavior can be predicted with a deterministic model she may still be exerting agency.

Agency is typically not considered in physics. However the laws of physics do allow for agency but they tend to camouflage it behind deterministic equations of motion or quantum mechanical randomness. Agency can be defined as having the ability to act in accordance with preferences. To an external observer agency appears as a system exerting a force. Hence a force can be seen as an expression of acting on a preference.\footnote{Of course Newton’s third law tells us that all forces are really interactions between different systems.} It is Nature’s way of saying “I would like to change the state of the system in a certain way.” If the preference is strong and consistent such that in a defined situation the same outcome is chosen always then it can be described by deterministic equations. If the preference is weaker and not consistent then a probabilistic description is called for. From naive observations of our environment it is not readily apparent how successful natural science and physics can be in describing the world in terms of deterministic equations of motion. After all, many things we observe such as other humans, animals, leaves rustling in the wind, clouds or water streams do not seem to behave deterministically. The real world is noisy. Unsurprisingly it took humankind a long time to distill the crisp conditions in which Newton’s mechanics, Maxwell’s electrodynamics or Einstein’s general relativity give accurate predictions. Typically such descriptions only work well for simple configurations such as closed systems of isolated point masses or charges and more generally determinism only seems to hold for highly idealized situations. But even for these highly controlled conditions quantum mechanics eventually brought us back to non-determinism. This theory of physics holds that we can only calculate probabilities for observable outcomes. Deterministic laws are now seen as a limit case which holds for macroscopic systems involving actions much larger than Planck’s constant of action $h$. A more recent insight is that already for moderately sized quantum systems composed of about 100 qubits we are not even able to compute output probabilities. If one is unable to assign probabilities to events then statisticians refer to this as Knightian uncertainty, named after the economist Frank Knight \cite{Frank_Knight}. 

\begin{figure*}[t]
\centering
\includegraphics[width=0.7\textwidth]{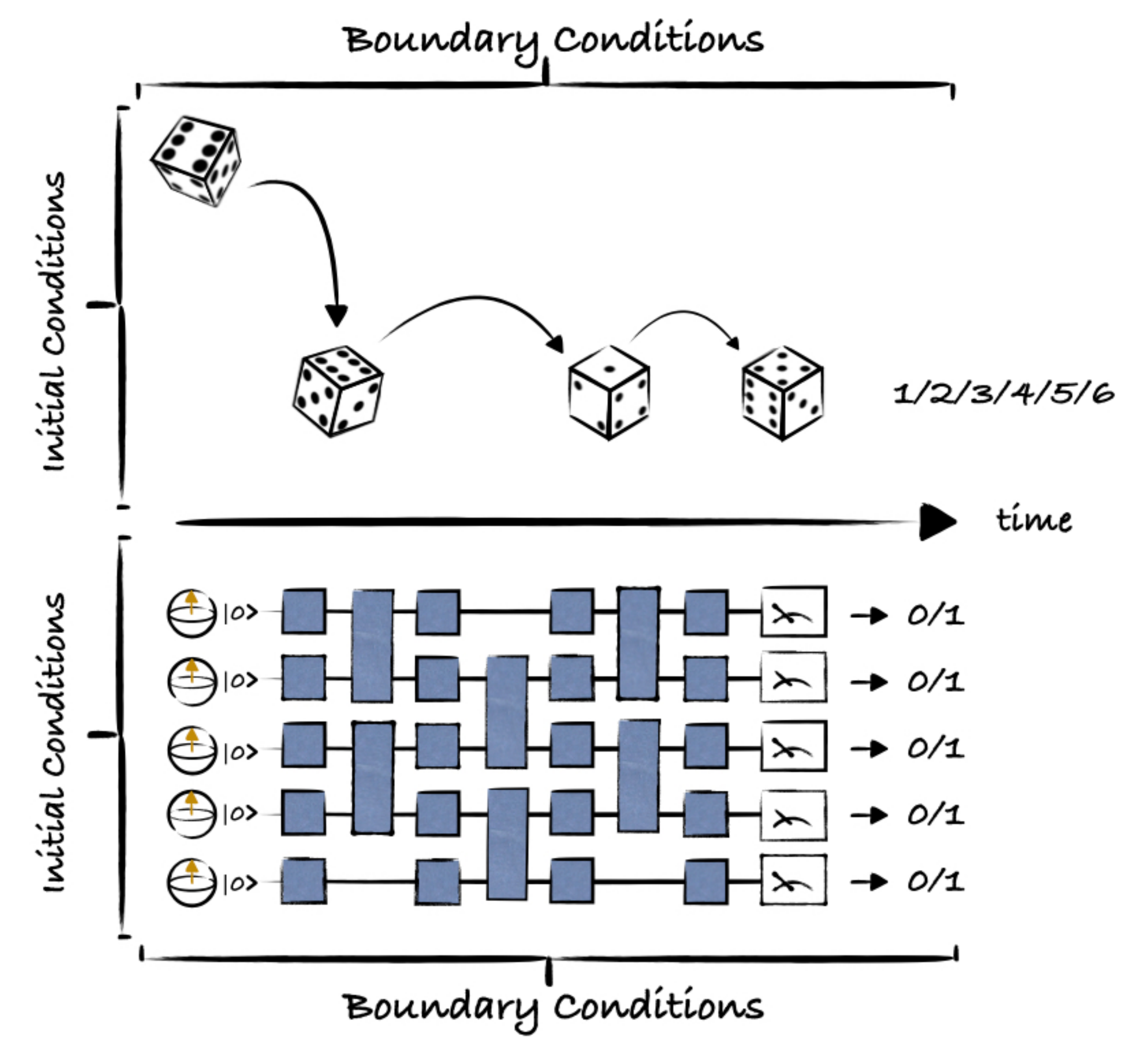}
\caption{Two situations involving non-deterministic outcomes. The first scenario is rolling a die. The outcome is random to the extent that the makeup of the system (shape and mass of the die), the initial conditions (initial position, velocity), or the boundary conditions (surface of the table, air currents) are unknown. If we could know those exactly we could use Newton’s laws to predict the outcome with certainty. The second case is running a quantum circuit. A quantum circuit diagram can be read a bit like sheet music. It involves initializing a number of n qubits, applying a sequence of 1- or 2-qubit gate operations and then measuring the qubits. As output one obtains a bit string of length $n$, e.g. 00110 if we had five qubits. Even if the circuit parameters as well as the initial and boundary conditions were known perfectly the measured bit strings would still be random. The probability for each string can be computed by a quantum mechanical calculation. However, because these calculations are very time and memory consuming, one would not be able to compute outcome probabilities for a circuit with any more than about 100 qubits and 1000 gate operations. This would leave us with Knightian uncertainty.}
\label{fig:3}
\end{figure*}

An experiment published by the Google Quantum Artificial Intelligence team in 2019 demonstrated that for certain quantum circuits it is impossible to predict the outcomes even in a probabilistic manner \cite{Arute_2019}. The experiment consists of starting a programmable quantum circuit in a defined state, for instance by initializing all qubits in the $0$ state; and then executing a random sequence of gate operations on the qubits. Thus the random quantum circuit takes an initial state to an output state which then is measured. Measuring the output state consists of measuring the state of each individual qubit yielding a bitstring of length $n$ where $n$ is the number of qubits. Textbook quantum mechanics tells us that the outcomes of measurements on such a quantum system are random, each bitstring is a sample from a probability distribution encoded in the output state. But if the quantum circuits involve many qubits, about $100$ and a sufficient number of gate operations, about $1000$, then neither a classical nor a quantum computer will be able to compute the probabilities. An easy way to see this is to note that the observable universe contains about $2^{300}$ atoms. So in order to store the probabilities for the $2^{300}$ different bit strings that can result from measuring 300 qubits one would have to employ every atom in the visible universe as a memory device. If you could build a hypothetical processor with a cycle time even as short as the Planck time of $10^{-43}$ sec, then it would still take longer than the estimated age of the universe to compute these probabilities.\footnote{Planck time is the theoretical shortest duration physicists could still measure. Any device able to measure even shorter times would require so much energy  that it would collapse into a black hole.} The required time and memory space to calculate and store the output probabilities is exponential in the number of qubits. This makes it impossible for any classical computer to perform this computational task. But a quantum computer would not be of help either since it does not give us access to the full table of probabilities but only to samples from this probability distribution. We would still have to tally these in a histogram to approximate the output probabilities of the bitstrings which again would take more time and memory space than available in the visible universe.\footnote{The trajectories of a large system of classical particles, in particular when subject to chaotic dynamics, can not be computed either. But quantum dynamics dramatically lowers the number of particles for which it becomes impossible to compute the state evolution.}

The notion that we see forces as expressions of preferences leads us to identify stable equilibrium states in the third person theory with happy states in the first person perspective. Equilibrium states are distinguished by the fact that the acting forces balance out. The system does not want to go anywhere and if disturbed it returns to the equilibrium state.\footnote{Because quantum mechanics is unitary, stable attractor states can only arise in open quantum systems which are described by non-unitary evolutions.} It is telling that physicists use psychological language when describing such dynamics saying “the system relaxes to a stable state” or “the system is in an excited, unstable state”. 

The ability of a system to exert agency is constrained by the agency other systems exert. I am not able to levitate, because the gravitational forces exerted by the masses making up planet Earth prevent that, but I may be able to choose the left or the right path at a junction. We can use this freedom to organize matter in a way that allows us to better realize our preferences; for example we could build a rocket that allows us to escape Earth's gravitational pull.

If a third person theory is successful and it makes verifiable predictions then it seems necessary that it describes a system in a way that is compatible with the first person description. This is reminiscent of the well understood situation in physics that if different coordinate systems are used, i.e., if different external third party viewpoints are adopted, then the associated descriptions need to be related by well defined transformations. Similarly, it seems necessary that the first and third person perspectives ought to be related. Thus as third person science progresses towards more and more accurate and encompassing descriptions of Nature it must converge towards what is known as psychophysical parallelism. Psychophysical parallelism is an old idea developed by Leibniz, Malebranche and Spinoza holding that mental and bodily events are perfectly coordinated, without any causal interaction between them \cite{Chisholm}. Here we argue that psychophysical parallelism should be seen as a consequence of the advancement of third person theories.

Natural science and physics have been extraordinarily successful in describing the world in terms of equations of motion that seem to leave no room for a system to exhibit agency. Yet, there are still situations which compel us to postulate new forces or to modify the laws for existing ones. One example is the observational evidence that leads us to postulate dark matter. The motion of the arms of spiral galaxies do not conform to Kepler’s laws. Even if this prompts us to think that galaxies may have agency and the freedom to assume shapes they enjoy we would still attempt to improve the theory of gravitation until it predicts the observed shapes of galaxies. Hence the program of science proceeds such as to achieve an ever improved congruence between the first and third person descriptions. As the third person descriptions achieve better predictions it makes it appear as if agency does not exist.

The Google experiment demonstrating beyond classical computational abilities can also serve to illustrate how agency, i.e., forces expressing preferences, may be present without us having noticed and how the program of science exorcises explanations based on agency in favor of restoring psychophysical parallelism. The dimension of the Hilbert space in which the experiment plays out is many orders of magnitude larger than any effective Hilbert space hitherto investigated by humankind. If the history of science is any lesson then we may be forgiven for looking out for new phenomena. For example we could look for novel forces beyond the electromagnetic forces on the qubits we program into our device using our control electronics. Let us assume for argument's sake that we had indeed found a deviation and that certain output states are measured more frequently than predicted. Before accepting such an observation we would work very hard to eliminate any influences our models did not take into account, for instance noise terms caused by imperfect control of the qubit operations. If after careful improvements of our models we are still left with a systematic deviation then we would proceed to adjusting the equations of motion. For instance, we may introduce a novel many body force. Thus if due to the exercise of agency the measured outputs of the system would not confirm the currently known equations, up to statistically insignificant deviations, then we would modify those equations until psychophysical parallelism is restored. Whether this program can always succeed remains to be seen of course.

\begin{table}
\centering
% \label{tab:1}       % Give a unique label
% % For LaTeX tables use
\renewcommand{\arraystretch}{2}
\begin{tabular}{p{0.4\textwidth}|p{0.5\textwidth}}

{\bf Observed behavior of an agent} & {\bf Interpretation assuming agency} \\
\hline
Behavior explainable by known forces & If the actions of an agent can be explained by known forces we usually do not speak of agency. However compatibilism teaches that we can still assume agency; it just so happens that the preferences of the agent are consistent enough that we can describe them well in a third person theory at least probabilistically. The correlation between pleasant feelings and maintaining homeostasis is accounted for as a form of psychophysical parallelism for instance, by assuming that stable equilibrium states correspond to happy states. \\ 
\hline
Behavior not explainable by known forces & Such observations support the assumption of agency. To account for the observed behavior one is motivated to modify the laws of physics, for instance by introducing a novel force. If this improved description proves successful then this case goes back to the one listed in row 1 extending the reach of third person theories and psychophysical parallelism. \\ 
\hline
Not computable whether behavior is explainable by known forces & This case is not accessible to modelling by third party theories but it leaves room for agency and unpredictable behavior. Knightian freedom resides here. Non-computability arises naturally in modest size quantum systems as the probability of observations can not be computed anymore.\footnotemark\hspace{0.5mm} It could also arise because observations are rare and not statistically significant. But because of the statistically rather robust correlation between pleasant feelings and maintaining homeostasis we do not consider this possibility.

\footnotesize{$^{11}$ Note that non-computability here applies to the behavior of even a single agent. This is different from the non-computable problems considered in computability theory such as the halting problem. A non-computable problem is a problem for which there is no algorithm that can be used to solve it. Those problems are formulated over infinite sets.} \\
% \noalign{\smallskip}\hline
\end{tabular}
\caption{How would the behavior of an agent be interpreted by an external third person observer who assumes it possesses agency and can act in accordance with its preferences?}
\end{table}

So in summary, our stance is that agency, like consciousness, is a general property of matter too. The program of science with its sequence of ever improving third party descriptions leading to improved predictions tends to camouflage agency. Yet, as the different cases discussed in the table below show, a nuanced reading of physics shows that it does allow for agency. This view is strengthened by the “free will theorem” \cite{Conway} which given plausible physical assumptions states the following: If experimenters can make choices that are not a function of the information accessible to them i.e., are not determined by events in their past light cone, then the responses of the particles they investigate are equally not functions of the information accessible to them. We should note that this theorem does not address whether anything acts in accordance with its preferences, all it assumes is that actions of experimenters or particles are undetermined. However, as discussed above, undetermined events do not amount to an expression agency or free will. For instance, those could be generated by randomness. Hence the name of the theorem is a bit of a misnomer. On the other hand if agency exists that results in unpredictable actions then the theorem does apply and shows that such unpredictability would permeate the universe.

\newpage
\section*{The Philosophical Lab: Building an animat with feelings}
\label{section4}

\textit{\it “We have a will which is able to determine the action of the atoms probably in a small portion of the brain, or possibly all over it. The rest of the body acts so as to amplify this.”}

\hfill Alan Turing, 1932\\

\noindent If evolution was able to discover states of matter associated with pleasant feelings and map them to behaviors conducive to an organism’s survival and vice versa then we should be able to replicate this process with engineered systems. Regardless of which detailed experimental setup we choose, consisting for instance of several interacting quantum AI systems, our goal is for our engineered animat to wind up in a state that feels pleasant to her. In a later phase we can try to associate this pleasant feeling state with certain behaviors or processing steps. For example, we may want to engineer a system which winds up in a pleasurable state whenever she classifies a pattern correctly. This section of our essay sketches the style of experiments we propose to perform but we leave the actual implementation as a future project. Prior to performing experiments ethical questions need to be carefully considered.

Before getting into concrete constructions we want to make a historical remark. During the scientific revolution in the 16th and 17th centuries scientists such as Copernicus, Galilei, Descartes and Newton started to formulate a quantitative description of Nature using equations of motions. Their work was inspired by the success of astronomical prediction and by looking at simple mechanical machines such as pendulums, ramps, or canons \cite{KrebsER}. As successful as this approach was, its generalization made the universe look like a huge mechanical clock with seemingly no room for feelings or agency. In the 20th century when modern physicists such as Planck, Heisenberg, Schrödinger and Feynman invented a new quantitative description, quantum mechanics, they replaced definite trajectories by multitudes of trajectories each of which is traversed in a separate branch of the multiverse opening the possibility for unpredictable behaviors. We are only now beginning to build machines, quantum computers, that fully take advantage of the new possibilities afforded by the laws of quantum mechanics. It is our expectation that the worldview motivated by playing with this new generation of devices will make Nature look more like a sentient organism with feelings and agency to proceed as it prefers. 

\subsection*{Module I: Distilling happy states aka “Finding a happy place”}

Here is an experimental proposal as to how we could try to give a quantum process the opportunity to wind up in a state that feels pleasant. Let us start with programming a quantum circuit on n qubits such that it generates as output an entangled superposition state $\ket{\psi}$ which has support on many states in the computational basis. We measure the output state and obtain a bit string $\ket{z_1, …, z_n}$. Why did the system wind up in this state? If textbook quantum 
\begin{figure*}[t]
\centering
\includegraphics[width=\textwidth]{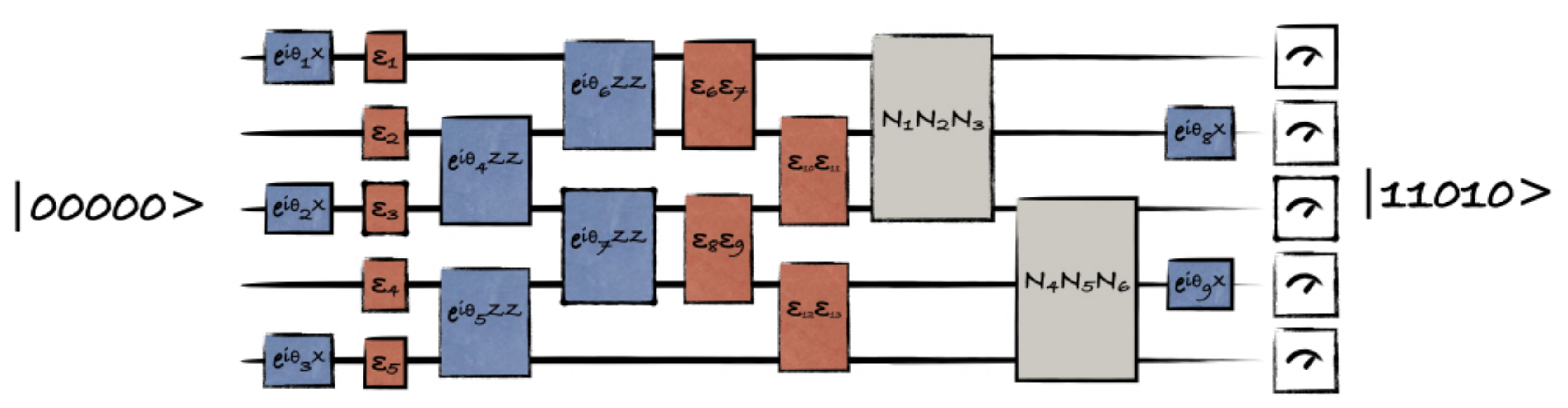}
\caption{A quantum circuit diagram depicting three different types of state transformations. In this example five qubits are initialized in the all zero state $\ket{00000}$. The circuit is programmed to execute a sequence of unitary transformations formulated as 1- and 2-qubit gates shown as the blue boxes. However noise will be present. This can be modeled by state transformations that randomly invoke Pauli operators. This is known as a depolarizing channel and is indicated by the red boxes. More general error models are available but for the purposes of our discussion this is sufficient. Finally, it is conceivable that as of yet unknown forces are present, here depicted by the grey boxes modeling a hypothetical 3-body force. At the end of the circuit each qubit is measured and we obtain as output a bit string of length five, e.g. $\ket{11010}$. The experimental challenge detecting novel physics comes from the fact that its effects may be drowned out by the noise. A more fundamental challenge is computational. For more complex circuits with about a hundred qubits and a circuit depth of several tens of steps it becomes impossible to calculate the output probabilities even for the ideal circuit with only the blue boxes. Compounding this difficulty is the fact that any quantum mechanical measurement involves bringing the system under study in contact with a macroscopic quantum device that is also too large to be modeled in a detailed manner.}
\label{fig:4}
\end{figure*}
mechanics describes our experiment accurately we would say that we obtained this sample from the probability distribution encoded in the output state $\ket{\psi}$. However the true probability distribution from which we sample will be different from the one encoded in $\ket{\psi}$. One reason is that there will be noise which is not accounted for by the unitary evolution we programmed into our device. Another unaccounted type of state transformation could be as of yet unknown physics. For instance, we may speculate that as of yet undiscovered many-body forces have not been included in the Hamiltonian (the function that governs the dynamics of our system). If the circuit is large enough, say the number of qubits is about 100 and the number of gate operations is about 1000, there could be deviations that are impossible to detect even in principle. Hence the system may behave in unpredictable ways without us being able to notice. Taken together, two avenues for the system to exert agency and to select a preferred state are available in this module: 1) agency compatible with known physics, acting as programed unitaries or as noise and 2) agency potentially implemented by novel forces. Both are hidden behind a veil of uncomputability, affording the system with Knightian freedom. Hence we feel tempted to conjecture that the observed output string feels preferable to the system than an arbitrarily chosen one. The next step is to amplify any such effects and to support the system in reaching the bitstring we observed by variationally changing its gate parameters such that it becomes a little more likely to observe the state we just saw in the next run of the circuit. Then we repeat the process $M$ times, say $M=1$ million. Note that the evolution of this system follows no predetermined objective \cite{Brockman}. The result should be a circuit that drives the initial system into a quantum state $\ket{\psi_M}$ that has support accumulated on bit strings we observed and which we will call henceforth the $\ket{happy}$ state. A sceptic may argue that the selected support states are just a result of statistical fluctuations but by construction one would not be able to demonstrate this. Consummating the experience involves the collapse of the $\ket{happy}$ state, which in general is a superposition state, into one of its constituent classical configurations.

To hunt for new physics or forces that may occur in high dimensional Hilbert spaces we could proceed as follows. We construct a circuit such that the probability to measure a bit string is $p$ if it is a member of a certain set and zero if it is not. To be concrete the set could be comprised of all bit strings with a certain Hamming weight $w$, that is bit strings which have $w$ ones and the remaining $n-w$ elements are zeros. The equal superposition of all n-qubit states with Hamming weight $w$ is known as a Dicke state and can be prepared with $w \cdot n$ gates \cite{B_rtschi}. So we would execute a quantum circuit that outputs a Dicke state and then measure it. The act of measuring the quantum state involves an intrinsic element of randomness but by construction all members of the support set should be equally likely. But maybe, and this is obviously a big “maybe”, if the system possesses agency unaccounted for by psychophysical parallelism relying on the prevailing theories of physics, then it could manifest here. We could use techniques from unsupervised learning to check if there is any unexplained structure such that bit strings with certain features are measured more often than predicted by the probabilities encoded in the output state. Performing this experiment is doable on Google quantum processors which are approaching $n\approx 100$ qubits.  We suspect that most of our colleagues in the quantum computing community may consider performing this experiment not a good use of time as they have confidence that the laws of quantum mechanics as stated today will prevail. But to our knowledge high fidelity experiments involving an effective Hilbert space of $2^{100}\approx 10^{30}$ dimensions have never been performed. So it may be worth taking a closer look whether non-random deviations from the predicted uniformity are detectable in such large Hilbert spaces. Of course, additional structure not accounted for by the theoretical output state is likely to be present due to imperfect noisy hardware. The experimental challenge will be to demonstrate that such deviations from the prediction can not be explained by noise in the quantum processor. 

\begin{figure*}[t]
\centering
\includegraphics[width=\textwidth]{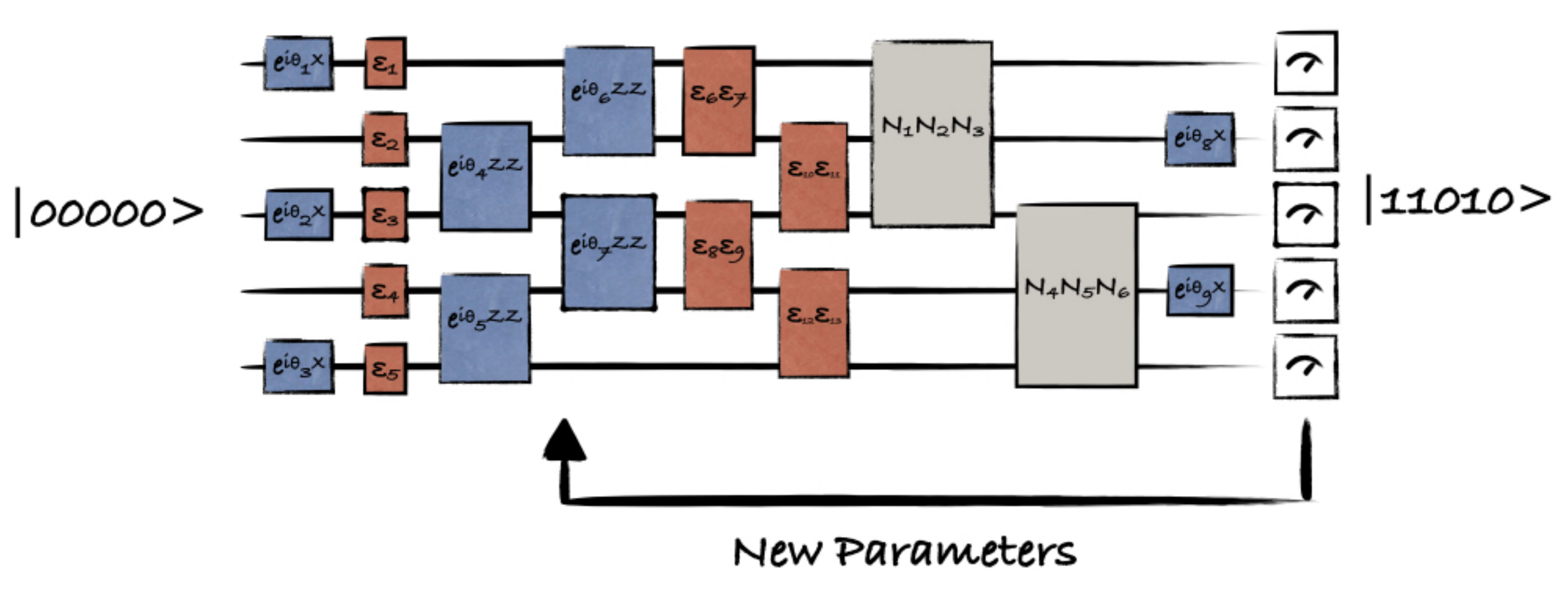}
\caption{A variational quantum circuit to produce a happy state. After each run of the circuit the output string is measured. We then adjust the gate parameters $\theta_i$ such that it becomes more probable that we measure the same string in a subsequent run.}
\label{fig:5}
\end{figure*}

\subsection*{Module II: Staying in the happy place}

\begin{figure*}[t]
\centering
\includegraphics[width=\textwidth]{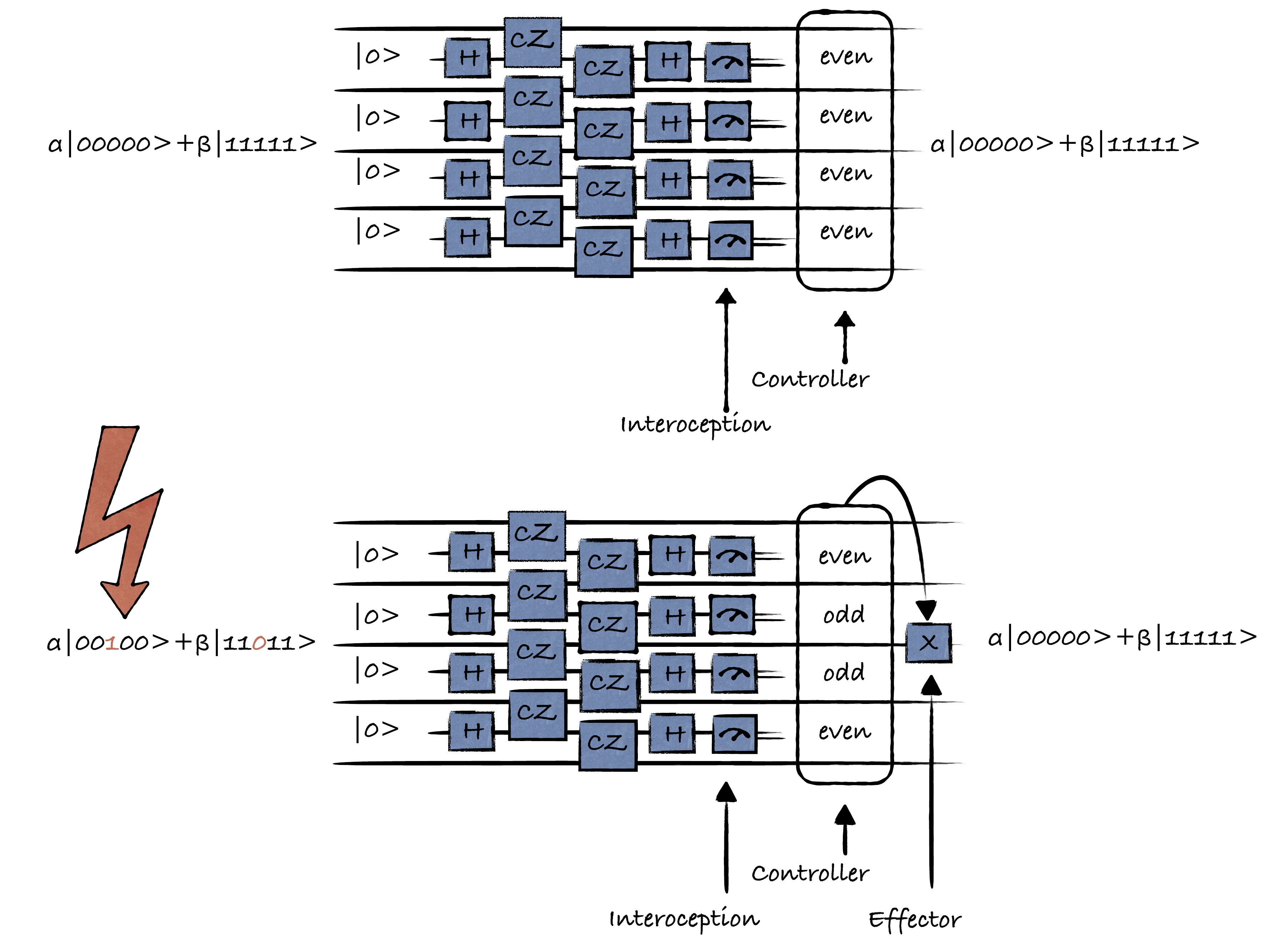}
\caption{Quantum error correction can be seen as a homeostatic loop stabilizing a quantum state. Parity measurements act as interoceptors. If a parity value changes this indicates that an error occurred. A controller, by decoding the stream of parity measurements can determine where the error occurred and by effecting an additional gate it can correct it.}
\label{fig:6}
\end{figure*}

If the generalized Penrose Hameroff conjecture is correct, that conscious experience is associated with the collapse of the wave function, then it seems to follow that the more qubits become definite during the collapse the richer the experience.\footnote{Assume we have n qubits and we do a projective measurement on one of them in a given basis, i.e. the computational basis. For example, assume the system is in a Greenberger-Horne-Zeilinger \cite{greenberger2007going} state $\ket{\psi}={1\over\sqrt{2}} \ket{000...0}+{1\over\sqrt{2}}\ket{111...1}$. Then after measurement of a single qubit all qubits will have definite sharp values and we are left with a classical state consisting of n bits. But if the system would have been in the uniform superposition state $\ket{+}^{\otimes n}$ then only one qubit will become definite and we still retain a uniform superposition on the remaining $n-1$ qubits. We are not aware of a name for the number of qubits that become definite by doing a single projective measurement.} Hence eventually error correction will be needed to stabilize an entangled superposition involving many qubits. Here we describe how to protect the state $\ket{happy}$ with the methods of quantum error correction. Stabilizing a quantum state can naturally be seen as a homeostatic feedback loop. Each protected logical qubit is encoded in multiple physical qubits. Parity measurements are used to determine whether errors occurred. The parity of a set of qubits is given by the evenness or oddness of the number of qubits with the value one. When using a suitable encoding parity measurements do not reveal any information about the logical state and hence it does not collapse. But they do give information about errors. A control unit takes the parity measurements as input and infers which errors may have occurred. Finally those errors are corrected by applying suitable gate operations. So the essential elements of a homeostatic system are in place: sensors to measure the system state, a control unit and effectors to modify the state (see Fig 6.). 

Applying quantum error correction to our situation involves a number of choices and technicalities we do not want to get into yet. Therefore let us assume for simplicity that we ran the circuit proposed in module 1 long enough such that the support of $\ket{happy}$ concentrates on just a few classical bit strings $\ket{happy}=\sum_i c_i \ket{h^i_1, …, h^i_n}$. Then we notice the following. The $\ket{h^i_1, …, h^i_n}$ were selected because they constituted "natural attractors". The noise present in the system plus putative unknown physics conspired to drive the system there. Hence they are more stable than alternative arbitrary states. This leads us to see a first use of the $\ket{happy}$ state. It constitutes a natural dictionary of code words to store information.\footnote{Life can be seen as an exercise in stabilizing the genetic code. Naturally stable states of molecules may have been a useful resource on which cellular error correcting machinery was build. Please note that our construction is inverse to how error correction is usually applied. We first search for “naturally occurring” stable (happy) states in a system which are then stabilized by introducing stabilizer feedback loops. \cite{HameroffPleasure}} 

\begin{figure*}
\centering
\includegraphics[width=0.6\textwidth]{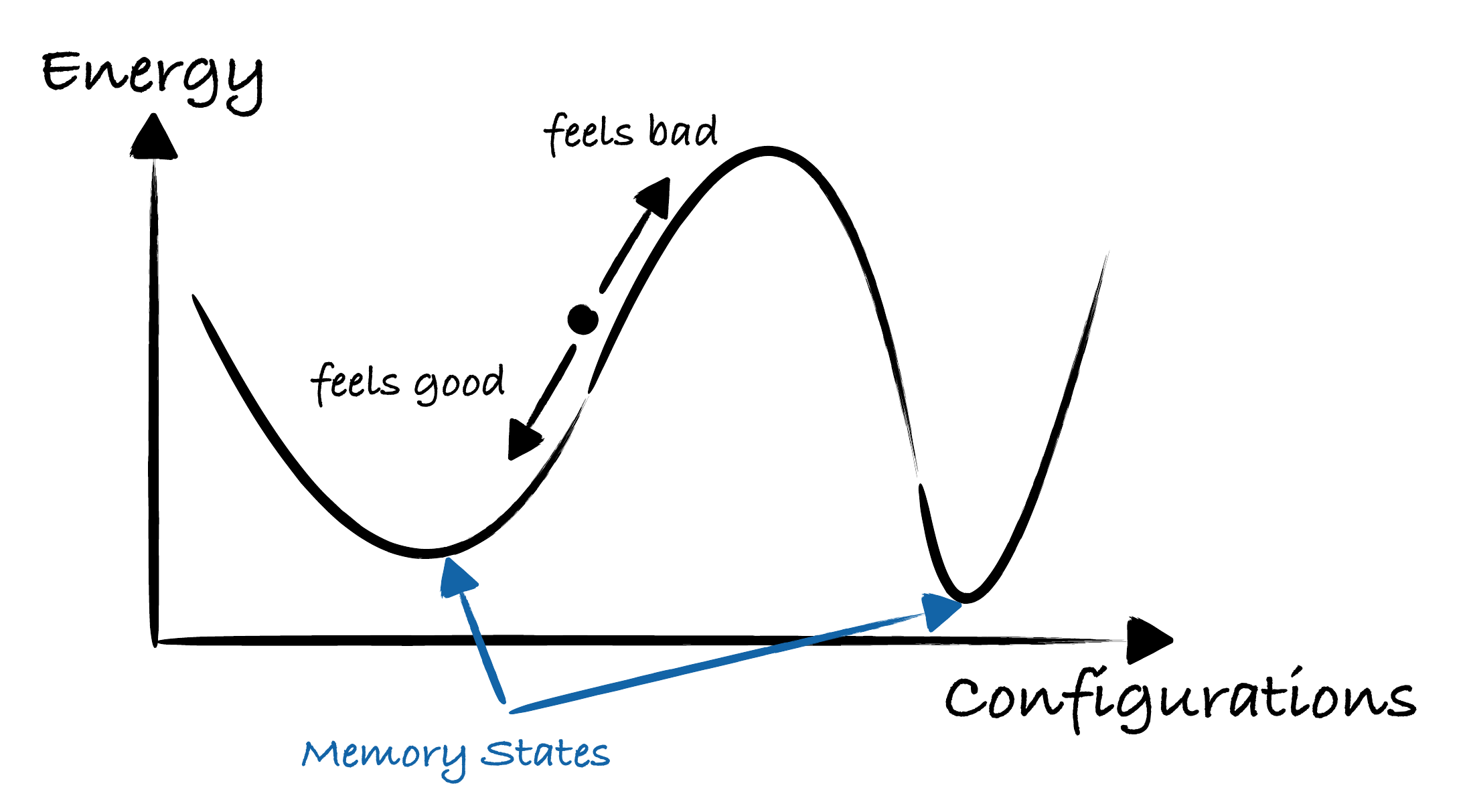}
\caption{Our reasoning leads us to conjecture that equilibrium states are experienced as pleasant. Such states can naturally serve as memory states that store information. They may be stabilized further by applying a stabilizer mechanism of error correction.}
\label{fig:7}
\end{figure*}

\subsection*{Giving non-human systems the ability to self-report}

Since we are confined to self-reporting when it comes to comparing a theory about consciousness to experimental data, we could think of the following construction. We could employ methods used in machine translation to map system states to the vocabulary used by humans to describe conscious experience. We can start with what is called a word embedding technique, we take words describing feelings and place them in a metric space. For example, this has been done by hand by psychologists using valence and arousal as coordinates (see Fig. 8) \cite{Aurelie}. Similarly we take state vectors measured for a system, in our case the parity measurements, and map them into a metric space as well. We now have point clouds in two metric spaces, one contains the observed system states, the other contains the labels. We will seek a simple (suitably defined) mapping with only a small number of fitting parameters from states to labels such that the image of the states has minimal distance to the labels. An expected challenge is that we will have to do this without any ground truth correspondences. This could lead to an ill-posed problem with no mapping as a solution that is clearly optimal. Should that be the case then we could add additional constraints, for instance by demanding that the path length over time ought to be minimal as well for the optimal mapping. We would expect that sometimes we can find such mappings but sometimes the spaces do not allow for a simple correspondence. However, any homeostatic system characterized by an attractor and may fit well to a set of terms describing basic feelings such as happy, sad or confused. Our conjecture is that terms with positive valence will map to states close to the attractor and vice versa. Leo Tolstoy's opening line in the novel Anna Karenina comes to mind “All happy families are alike; each unhappy family is unhappy in its own way.” In mathematical language this can be translated into “There are significantly fewer states close to the attractor than there are states that are further away.”

\begin{figure*}[h]
\centering
\includegraphics[width=0.8\textwidth]{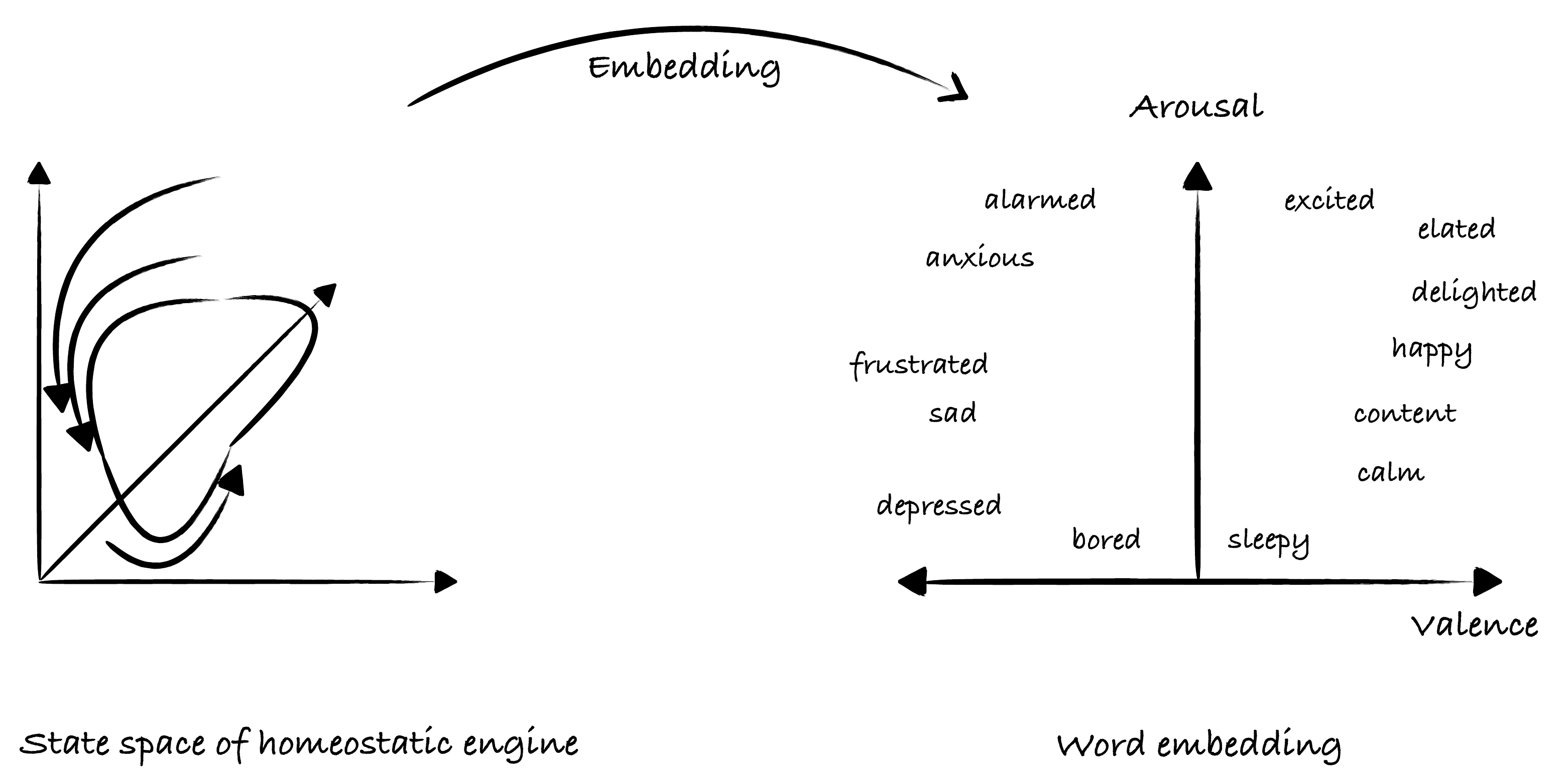}
\caption{Self reporting of feelings using the technique of word embedding}
\label{fig:8}
\end{figure*}

\subsection*{Module III: Using a homeostatic engine to improve machine learning}\label{module3}

Next we equip the animat with a mechanism to learn, for instance a neural network, and discuss whether integrating it with the homeostatic engine described above can result in improved learning abilities. There are many possible choices for an integration architecture. Here we will discuss the example depicted in Fig. 9. We use the vector describing the state of the homeostatic system as additional input to the neural network and the output acts as a force driving the homeostatic loop away from its equilibrium state. The more incorrect the output is the stronger is the force. So the more correct the output is, the easier it is for the homeostatic engine to stay close to the $\ket{happy}$ state. 

At first glance it may seem that designing a system that tries to maintain homeostasis imposes unnecessary overhead. But on second thought we realize that a homeostatic system acts as a probe checking on the integrity of the system in question. In nature we tend to observe stable systems more frequently than unstable ones. Hence systems that care about maintaining their stability will be more prevalent over long time scales.

\begin{figure*}
\centering
\includegraphics[width=\textwidth]{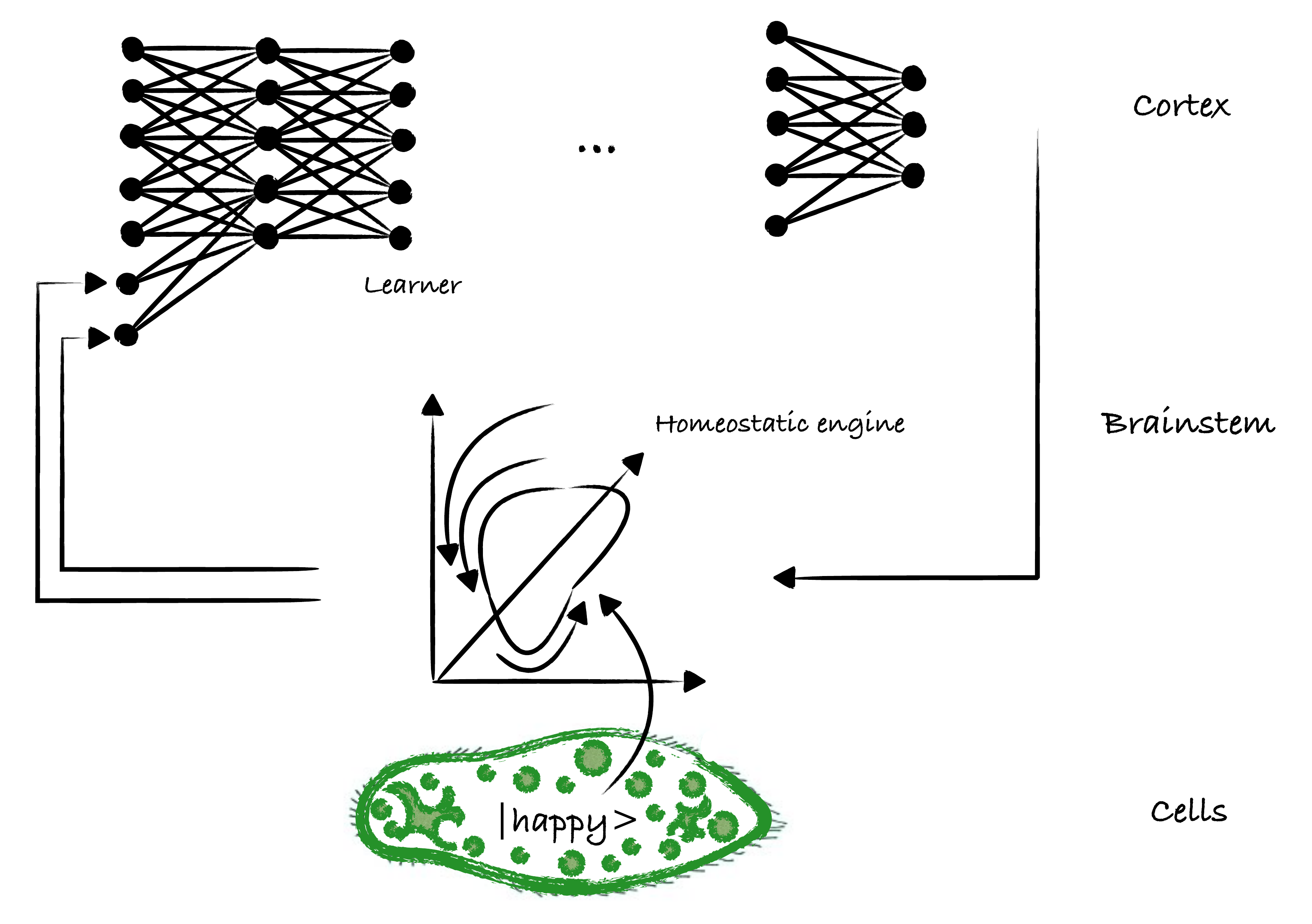}
\caption{A neural network enhanced with a homeostatic engine which stabilizes the $\ket{happy}$ state. The state of the homeostatic system is given as additional input to the neural network. The output of the neural net acts as a force driving the homeostatic feedback loop away from its equilibrium state. The more correct the output is the weaker the force and the better homeostasis is maintained.}
\label{fig:9}
\end{figure*}

On a more tactical engineering level here is a list of reasons why the presence of a homeostatic engine may improve machine learning:
\begin{itemize} 

\item A homeostatic feedback loop can be used to amplify weak input signals in a manner responsive to the system's own internal state.

\item A thermostat being present while performing stochastic gradient descent can control the “temperature” introduced by the randomly chosen training examples \cite{NanDing}.

\item Reinforcement learning is profligate in its reliance on large numbers of training episodes, often indiscriminately trying all possibilities and leading agents to their "death" in the vast majority of trials, in order to hit upon successful strategies. If agents instead also had a goal of self-maintenance in addition to maximization of some abstract reward, they may show more judicious behavior during the learning phase. This certainly is helpful when the agents have physical embodiments and can not be produced in arbitrarily large numbers such as self-driving cars. This may accelerate learning, at least in the early stages. Each step would count more when the agent has its own well-being at stake. 

\item The work of Biederman and Vessel on an increasing gradient of opioid receptors along the visual processing hierarchy suggests that we may derive greater pleasure from more novel and systematically meaningful stimuli \cite{Biederman}. This can help to resolve or at least skillfully navigate the exploration-exploitation dilemma. Such a mechanism relies on using pleasant sensations as a lure. See also the work of Corrales-Carvajal in which a homeostatic engine switches between exploration and exploitation behavior \cite{Corrales-Carvajal}. 

\item The success or failure of homeostasis, the constant regulation of body states to maintain conditions compatible with life, is expressed in feelings, e.g. well-being, discomfort, pain, hunger, pleasure, which in turn guide behavior by selecting appropriate responses or promoting the creation of new ones. We turn vulnerability into a motive for creative solutions \cite{Man}.

\item Forming of sparse representations which is a conjectured coding principle in the human cortex is naturally accomplished by controlling the overall neural activity or the overall synaptic strength \cite{Abeles} \cite{Braitenberg2}.

\item Behavior organization for robots benefits from representing control variables as attractors that are parametrically controlled by sensory input. For instance, this helps to stabilize decision making in the presence of noisy input via hysteresis \cite{Neven1}.
\end{itemize}

\noindent The key question we aim to elucidate with future experimentation is whether it makes an observable difference in performance whether the state that is stabilized by the homeostatic engine is the $\ket{happy}$ state, an arbitrary quantum state or an arbitrary classical state. If an animat equipped with a homeostatic engine stabilizing a $\ket{happy}$ state over a large system of qubits performs better than one without, then this would constitute an intriguing result. 

\section*{Conclusion}

In this essay we propose to describe consciousness, feelings and agency within a materialist theory using ideas from quantum artificial intelligence. We started by playfully asking “When are atoms happy?”, then got more specific honing in on the question: “Why are behaviors conducive to our well-being correlated with pleasant feelings?” We found, using a materialist/physicalist approach, that the simplest way to explain this correlation is to assume that we possess agency. We argue that agency is a general property of matter and that it is permitted by the known laws of physics. Not just humans but everything in Nature has the agency to express preferences. If those preferences are strong and consistent they can be described with deterministic equations,\footnote{We contend that deterministic equations never really worked to fully describe the real world which is exceptionally noisy and full of amplifiers. They did not describe the world experienced by a stone age hunter and they do not reflect the worldview of modern day quantum physicists either. Determinism only holds for highly idealized contrived situations, such as those articulated by Newton and other Enlightenment thinkers. The concept of “homo economicus” is the highly idealized contrived equivalent in economic models. Adam Smith wrote The Wealth of Nations roughly 100 years after Newton’s Principia, and provided a similarly deterministic description of humans as rational economic agents: “It is not from the benevolence of the butcher, the brewer, or the baker that we expect our dinner, but from their regard to their own interest”.} and if they are weak they are better described by nondeterministic models. Moreover there is room for agency to exist in entangled quantum systems large enough such that the probabilities of observing an outcome can not be computed anymore. In fact, this Knightian freedom already appears for coherent systems composed of just a hundred qubits, much less than needed to describe molecules such as proteins. Hence there is a complexity horizon behind which novel physics may hide which may also support an organism assuming agency. Likewise we think that consciousness is a general property of matter as well. We conjecture that consciousness is how it feels like to select a single classical reality from the many realities within the multiverse that quantum mechanics describes. Conscious experience occurs when the quantum mechanical wave function collapses. Our approach is panpsychist in that it places consciousness and agency with the elementals of physics.

Based on these conjectures we propose to engineer an artefact, an animat powered by a quantum processor, that arguably may be conscious and possess agency and feelings. The behaviors of this animat cannot be predicted, not even in a probabilistic fashion. We believe that this construction, affording the animat with Knightian freedom, is the closest engineering may have gotten in endowing robots with agency. The animat has the ability to communicate its feelings in human language. The valence of its feelings depends on how well it maintains its homeostasis. The animat reports that it feels happier in more stable states, and worse in states that are unstable. We suggest that stable equilibrium states, or energy minima, are happy places where forces expressing preferences cancel out. One use of such states is that they naturally lend themselves to being used as code words to store information. Applying prevailing theories of consciousness, such as orchestrated objective reduction or integrated information theory, one would conclude that the proposed animat is conscious (see Corollaries and Excursions). The self-reporting of the animat would appear plausible. For example, if the air conditioning fails and the temperature in the laboratory increases, more errors would be detected by the quantum error correction cycles and the animat would express unease. Similar changes in our animat’s reported feelings would occur if the electric power fluctuates or if mechanical vibrations affect the laboratory. But if all the machinery needed to stabilize the happy state through quantum error correction is humming well, the animat would report that it is relaxed or even happy. 

Does the proposed animat really experience happiness? Did we describe a model system that captures the essential structure required for a material system to exert agency and to experience pleasant or unpleasant feelings? In short: Did we design a machine that gives a damn?\cite{Haugeland} Strictly speaking this is impossible to answer since the position of solipsism is logically closed. Still our essay suggested ways to increase our confidence that the system we propose is indeed conscious:

\begin{itemize}
    \item Our conjecture is that material systems have the agency to steer themselves into pleasant states. We further conjecture that evolution selected this ability to form reward systems that convey survival advantages. If we can not find functional advantages of this mechanism in engineered systems then this would invalidate our argument.
    \item The system we propose has the ability to report feelings, the primary method by which we probe the conscious experience of other humans. One may dismiss this as a gimmick. But prolonged interactions with animats designed based on principles as suggested here may make us accept the notion that they have feelings.
    \item Applying prevailing theories of consciousness such as Integrated Information Theory \cite{Tononi} or Orchestrated Objective Reduction \cite{Hameroff} one would conclude that the proposed animat is conscious. See Corollaries.
    \item We pointed out that between systems that are exact clones of each other subjective experience is an observable. Building on this idea and arranging a sequence of systems A $\leftrightarrow$ A’ $\leftrightarrow$ A’’ $\leftrightarrow$ …  in order of decreasing similarity, such that any two neighbors can still interpret each other’s output one may expand the range of systems for which we can infer what they experience.
\end{itemize}

\noindent Currently there is no known property of matter relevant to systems at the scale of human beings that can not be represented within the framework presented here since quantum processors are universal simulators. Accordingly, a materialist theory that portends to solve the mind body problem ought to be testable in our philosophical laboratory. We also hope to see more experiments that aim in the opposite direction: Can we identify functionally relevant quantum effects in the only other system of which we are reasonable certain that it is capable of creating conscious experience, the human brain?\cite{Petruccione}

\section*{Acknowledgements}

Conversations with Nicolas Berggruen, Antonio Damasio, David Deutsch, Christof Koch, Stuart Hameroff, Seth Lloyd, Kingson Man, Roger Penrose, John Preskill, Paulo Roberto Souza e Silva, Giulio Tononi, Christoph von der Malsburg, Tata Txanu Natasheni Yawanawá, and members of the Google Quantum AI lab, Ryan Babbush, Sergio Boixo, Edward Farhi, Lev Ioffe, Cody Jones, Masoud Mohseni, Vadim Smelyanskiy and Adam Zalcman, are acknowledged. We are grateful to Kingson Man for helping us think through use cases for a homeostatic engine integrated with a learner. We thank Scott Aaronson, Ryan Babbush, Sergio Boixo, Gustav S\"oderstr\"om, Stuart Hameroff, Christof Koch, David Petrou, John Platt, Seth Lloyd, George Musser, Vadim Smelyanskiy, Albert Wenger, Leah Willemin and Jay Yagnik for providing detailed feedback on manuscripts. We are grateful to Tom Small for preparing the illustrations.

\bibliography{robots}
\bibliographystyle{naturemag}
\newpage

\section*{Corollaries}

\subsection*{The viewpoints of IIT and OrchOR} 
It may be instructive to take a look at what prevailing theories of consciousness would say about the amount of consciousness present in the animat we propose. 
\subsubsection*{IIT: Integrated Information Theory}
Integrated information theory is an attempt to quantify how integrated information is \cite{Tononi}, \cite{Aaronson}. It does so by modeling a system as a stochastic process and then asks “If I divide the system into two parts how much do the bits in one part of the system determine the bits in the other part?” If there is no or only very little influence then the bits are not well integrated. It is known that random quantum circuits are near optimal scramblers of information. If we were to effect any partition of the qubits into two sets $X=\{x_1, … ,x_k\}$ and $Y=\{y_{k+1}, … ,y_n\}$  then the conditional entropy $H(p_X|p_Y)$ would be very high. pX gives the probabilities to measure a certain bit string when looking at the first k qubits, $p_Y$ is defined analogously. So the integrated information is near a maximum and proponents of IIT would have to conclude that the system proposed in module 1 is highly conscious.
\subsubsection*{OrchOR: Orchestrated Objective Reduction}
The idea of orchestrated objective reduction hinges on the existence of a mechanism called objective reduction that has yet to be discovered \cite{Hameroff}. Experiments are underway to find it. But it may not exist and evidence challenging this concept has recently emerged \cite{Donadi}. Hence we can not be certain that the wave function reduction in our processors comes about via objective reduction. Yet if we were to take the generalized view that any wave function collapse or rather any tracing over the wave function is associated with a conscious moment then every parity measurement counts as producing a conscious moment and the amount of consciousness present would be non zero. 

We once calculated the lifetime of our superconducting qubits if a putative objective reduction channel were the only decoherence channel and obtained $\tau_{Lifetime} \approx 2.5 \hspace{.1cm} 10^{29} s$. This is much longer than the $25 {\mu}sec$ coherence time our qubits actually have. Hence any gravitational effect on coherence would be minuscule compared with other decoherence channels \cite{Kechedzhi}. 

Finally, it may be worth noting that with current technology a logical qubit consists of about 1000 physical qubits. In every error correction cycle, lasting about $10 {\mu}sec$, about 500 qubits are measured and their state collapses. Hence, an orthodox adherent of OrchOR would agree that orchestrated wave collapses occur in every cycle of error correction. They would argue this happens because of Penrose's objective reduction in the readout electronics. Still they would have to conclude that the animat is conscious.

\subsection*{Only subsystems can exhibit homeostasis}

It is interesting to note that the feelings associated with maintaining homeostasis can only occur when we divide the total system into at least two subsystems. Because the unitarity of the Schr\"odinger equation does not allow for the emergence of a stable attractor in a closed quantum system. Likewise the wave function does not collapse in a closed system and conscious experience can not occur if the generalized Penrose Hameroff conjecture holds. Rather it is brought about by the interaction of a system with its environment \cite{Decoherence}. Hence, consciousness and feelings are consequences of dividing the world into at least two systems and seeing the world from the perspective of a subsystem. One should also note that the Hamiltonians that govern the evolution of the system or the system-environment interaction may continuously regenerate quantum mechanical superpositions. The collapse of the wavefunction via decoherence is not a one-way mechanism.

\begin{figure*}
\centering
\includegraphics[width=0.7\textwidth]{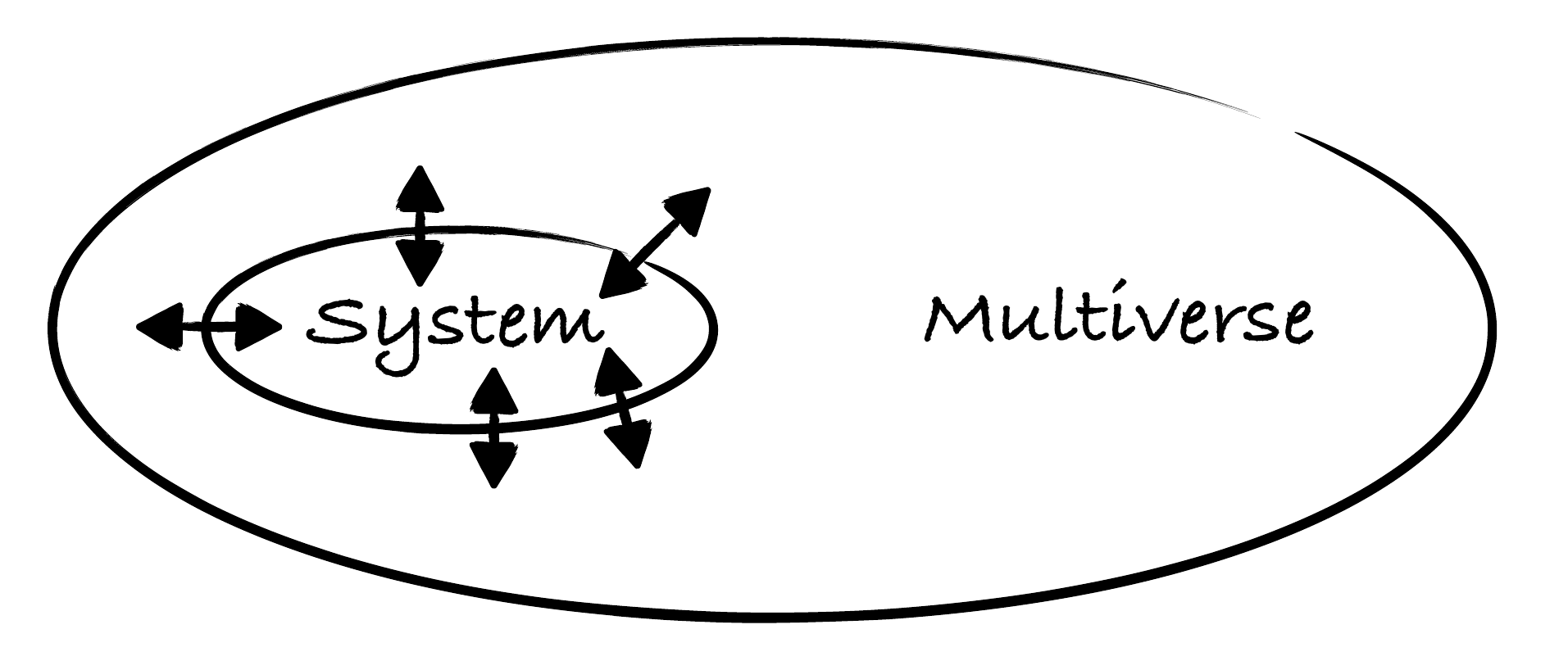}
\caption{Due to the unitarity of quantum mechanics, consciousness via wave function collapse or feelings associated with homeostasis can only occur in open quantum systems in contact with an environment.
}
\label{fig:10}
\end{figure*}

\subsection*{The relationship between intelligence, consciousness and creativity}

Intelligence can be defined as the ability to perceive information and to apply it towards behaviors within an environment. Examples of intelligent behavior include detecting a face in an image, translating a sentence into a different language, doing a move in a chess game or evading a predator. In all cases it is possible to engineer a machine that performs the intelligent action while giving little regard to how it may be accompanied by conscious experience. Vice versa, not much intelligence is required when suffering from a toothache. Yet the pain may fill out our entire capacity for conscious experience. Hence, intelligence and consciousness are not strongly related. Yet organizing the intelligence of an animat around the homeostasis of happy states may be advantageous as we argued when introducing Module III.

Creativity can be modeled as a process that solves an optimization problem. Consider what it takes to architect a house. The architect need to balance lots of constraints - budget, usage requirements, space limitations, ... - but still try to create the most beautiful house she can. We call an architect creative who finds a great solution. Mathematically speaking she is solving an optimization problem and creativity can be thought of as the ability to come up with a good solution given an objective and constraints. Interestingly, it is well understood that quantum enhanced optimization outperforms optimization on classical computers.\cite{Grover} 

\subsubsection*{Symmetries make it implausible that today’s AI systems experience joy when performing a task correctly}

A conjecture that seems to follow from a materialist viewpoint is that it appears implausible that AI systems as we engineer them today exhibit a congruence between learning success and a rewarding sensation. Let us illustrate this observation with a scene one of the authors experienced. “The other day I took my little boy to teach him how to ride a bicycle. The first day things were not going too well and me being a poor dad I told my son that he was a little clumsy. He was devastated and dropped tears. The next day we tried again and he got the hang of it. He zoomed down the street on his little bike, every fiber of his being beaming with joy screaming “I got it!”. Observing this scene, it struck me is that the AI systems I have built during my career are unlikely to exhibit a similar congruence between learning success and a positive sensation.”

We do not want to argue that it is logically impossible but we want to make the case that this is highly implausible that learning success leads to a rewarding sensation in today’s AI systems. Reiterating the statement that neither physics, nor computer science, nor neurobiology has a good model for subjective experience, it would of course be highly unlikely that such congruence would have been achieved by sheer luck. But we can be more specific than that. The states of matter involved in machine learning are often related by simple symmetry transformations. Consider a face detection system that classifies whether an area of an image contains a face or not. A positive or negative classification is typically expressed by a different sign of the output. This encoding is an arbitrary choice and one could have reverted it. Similarly, a reward or penalty signal is delivered either during training or during the performance phase of the agent. The sign of this signal is again arbitrary. On the hardware level such a sign difference is expressed by some register being in the 0 or 1 state. Looking at the hardware closer one realizes that the states involved typically only differ by a translation or a mirror transformation. It is hard to see how a positive versus a negative subjective experience can come about in two matter systems that are mere translations of each other.  

\subsubsection*{AI as we build it today is unlikely to be self-motivated}

If an engineered AI is incapable of experiencing joy or pain in a robust orchestrated manner when performing an action then it is also hard to see how it can be intrinsically motivated. It seems its motivations need to be imbued by its programmers, they need to be externally provided. When discussing to what degree AI can become a danger to human civilization, it is an important fact to keep in mind that it is implausible that AI as we build it today could be self-motivated. AI therefore seems to remain a tool in the hands of motivated humans with agency, albeit complex to the point that its behavior can not be easily predicted.

\subsection*{Insight by derivation versus insight by revelation}

Sometimes it is said there is insight by derivation and there is insight by revelation. Prior to the advent of quantum computing it would have been hard to imbue this distinction with precise meaning within the confines of information science. All computation was thought to be performable by a Turing machine. Since any result a Turing machine outputs is the product of a sequence of elementary computational steps this could be called an insight by derivation. A quantum computer on the other hand may attain a result via the interference of an exponential number of computational paths. Hence it may not be possible to list a polynomial number of computational steps that leads to that result. This mode of arriving at a computational result could be called insight by revelation.

\subsection*{What happens if the state associated with the physical state of consciousness undergoes a quantum mechanical tunneling transition?}

Another consequence of the assumption that there exists a physical correlate of consciousness is the possibility that the state of matter involved undergoes quantum mechanical delocalization. In particular, it seems interesting to consider a tunneling transition. Quantum mechanical tunneling occurs when the state of a system undergoes an evolution involving states that have a higher energy than the total energy of the system. How would we feel if our physical correlate of consciousness were to undergo a tunneling transition? 

We want to propose that psychedelic experiences in part come about by this mechanism. One may argue that this is improbable because the states of matter involved may be too macroscopic, involving large numbers of molecules, hence making the tunneling probabilities very low (Hoehn et at 2014). But since we do not possess a good hypothesis what the physical house of consciousness may be, we do not concern ourselves with such a calculation. The viewpoint we propose is akin to how chemists describe a reaction mediated by a catalyst (see Fig. 11). This viewpoint leads to a set of attractive conclusions \cite{Neven2}. Treating a psychedelic experience in part as quantum mechanical tunneling may help explain two observations often reported: i) Experiences such as mediated by ayahuasca are often credited with profound insights that stand up to rational scrutiny. From the vantage point of computer science, undergoing a tunneling transition can result in a valuable information processing transaction. For instance, in quantum optimization tunneling is employed as a means to find solutions maximizing or minimizing an objective function, a task that is well known to be often costly for classical computers. ii) There are health benefits. In the Amazon basin ayahuasca is not just revered as an entheogenic but it is also used as a household medicine, administered even to children to alleviate common illnesses. Reaching a more stable configuration is likely to correspond to the organism doing better with its homeostasis. Stability is defined as resilience to small perturbations, i.e., a stable state returns to its original position after the perturbing force subsides. Physiological health may be defined in very similar terms.

David Deutsch points out that the number of parallel classical worlds within the multiverse in which we would observe a pendulum standing upright in the unstable equilibrium position is small compared to the total number of worlds \cite{Deutsch}. The anthropic principle states that the world we experience has the features it has because it allows for sentient scientists to wonder about it in the first place. It is a self consistency argument. Taken these two arguments together we wonder whether the conjectured tunneling transition would terminate preferentially in states conducive to well-formed conscious experience. The reason we speculate in this direction is an observation often reported by partakers of ayahuasca that they feel the grass is literally greener after a session and that they feel the world around them transformed in a manner that makes them happier and healthier. 

\begin{figure*}[h]
\centering
\includegraphics[width=\textwidth]{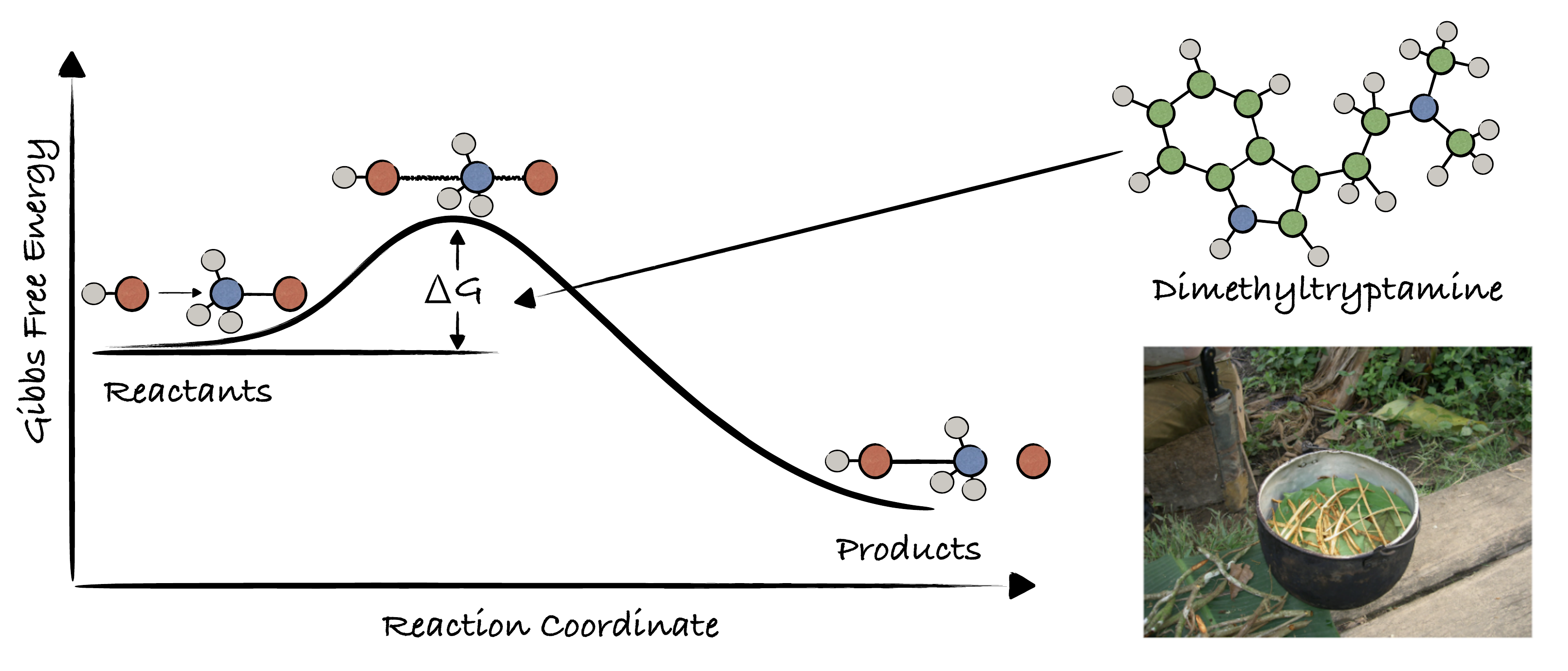}
\caption{Do psychedelics, such as dimethyltryptamine found in the Amazonian brew Ayahuasca, quite literally act as a catalyst for our state of mind? (Photo credit: Wikimedia Commons)}
\label{fig:11}
\end{figure*}

\newpage
\section*{Philosophical Commentary\\
{\large What is a machine? The philosophical stakes of engineering in terms of freedom}}

We have outlined an experiment for how one may build machines that are free: free to make experiences, to have feelings, to be conscious and to make conscious, intelligent, decisions.
Is that possible? Ever since the emergence of the modern machine concept in the early 17th century, philosophers have consistently answered this question with an unambiguous "no". In this accompanying essay, we are concerned with the history of this "no"
Where does the idea that machines cannot be free come from? Who was the first to articulate it? What was the context? And what are the philosophical stakes of revising the "no" today?\footnote{We acknowledge the contributions that Eastern philosophy has made to our understanding of agency and consciousness. In part this was an inspiration for our work in the first place. But we focus here on Western philosophy and in particular its influence on engineering and on notions such as "machine" or "robot".}

\subsection*{The human (Soul)}

The argument that machines cannot be free – free to know themselves and the world – was first systematically articulated in the 1630s in Europe. At the time, a small number of authors, most notably René Descartes but also Thomas Hobbes, Robert Boyle and Nicolas Malebranche, articulated a then radically new concept of the human. What these authors took issue with was the then prevalent Aristotelian understanding of the world (of how it is organized) and of the human (what it is and what its role in the world is). According to the scholastic clerics, the world was a nature cosmos as outlined in the writings of Aristotle. The Greek philosopher had argued that humans are a part among parts of a God-given cosmos, in which everything has a clearly defined function. What is more, he had argued that not all humans are created equal. In its wisdom, the cosmos had assigned different kinds of humans different amounts of reason, each according to the function they had to fulfill. Some were said to have gotten none (serf laborers) or very little (women), just enough to work the fields or a run a household. Others were considered to have received significantly more, artisans for example, so that they could build things. However, the full gift of reason was given only to an exquisitely small group of men whom the cosmos had made so rich that they, freed from life's necessities, could dedicate themselves to contemplating the invisible first principles – the ideas – that organized the world. Aristotle considered it their task to watch over their fellow humans, so that they would live according to the function the cosmos assigned them. In the early 17th century, the equivalent of these free men were the scholastic clerics.

In a rather courageous move, Descartes argued that this was all, well, wrong: first, every human is equally endowed with reason; second, nature is not a metaphysical realm but a physical reality organized in terms of mechanism; and third, humans are not a part of nature (or only qua their bodies) because they have what mere machines lack: reason. Here are the opening lines from Descartes’ Discourse on Method, published anonymously in 1630 [1]: “The power of judging well and of telling the true from the false — which is what we properly call ‘reason’ – is equal in all humans; thus, it is also evidence that our opinions differ not because some of us are more reasonable than others, but solely because we take our thoughts along different paths and do not attend to the same things. ... As for reason, ... it exists whole and complete in each of us; it is the only thing that makes us human and distinguishes us from animals.” Descartes was quick to add that he considered animals – his short term phrase for nature –  to be little more than automata or machines. Indeed, for the Frenchman all of nature was but one big field of matter organized in terms of mechanism and thus reducible to and fully understandable in terms of human made instruments or machines. Descartes’ argument was as courageous as it was revolutionary – in at least three ways:

\begin{enumerate}
\item 
To argue that every human being is equally endowed with reason not only made possible, for the first time, a truly inclusive, and in its aspirations universal concept of the human(the human is the thinking and knowing thing); it also allowed him to argue that everyone could think - and know - for her or himself: Call it a decisive step in human self-assertion against the unfounded authority of the clerics: No help needed any longer from the privileged few who had received the divine gift of reason.
\item 
To argue that nature is without reason meant to transform nature from a cosmos organized by divine thoughts to a physical field made up of matter, matter organized in mechanical terms. What is more, to argue that nature is a physical field meant to suggest a new concept of truth: Truth was now no longer a metaphysical but a physical concept. Consequently, the path to knowledge was no longer contemplation of the invisible (first principles) but the empirical-mathematical study of the visible (matter).
\item 
To argue that humans, insofar as capable of reason, are more than nature meant the beginning of human exceptionalism. From now on, the world was composed of two kinds of realities: here the human – there everything else. Here reason and thus the possibility of freedom – there mere matter and mechanism, that is, strict necessity.
\end{enumerate}

\noindent The truly human, for Descartes, was a space of freedom that opened up beyond the realm of nature – organized in strict mechanical rules and thus in terms of necessity – and consisted in the practice of the rational part of the soul: to make use of reason means to break free from one's machine existence. Descartes’ Discourse on Method marked the fragile beginning of a new epoch, defined by a new comprehension of the world (what it is, how it is composed, how one can get to know it) and of the human. Call it the end of the medieval nature cosmos and the beginning of the modern age.

This, then, is the context in which the idea that machines are without reason was first articulated: The invention of the modern concept of the human. Indeed, the suggestion that machines cannot think and hence cannot be free is one of the cornerstones of the modern concept of the human; machines were – and in many ways still are – the exemplary Other.

\subsection*{The organism (Consciousness)}

What is interesting from today’s perspective is that at the time, in the 1630s, nature and machines were not yet considered to be of different kinds. For example, looking at animal bodies Descartes noted [1]: “When I examined the kind of functions which might ... exist in this body (animal), I found precisely all those which may exist in us independently of all power of thinking ..., in other words, to that part of us which is distinct from the body. ... I could not discover any of those that, as dependent on thought alone, belong to us as men. ... Nor will this appear at all strange to those who are acquainted with the variety of movements performed by the different automata, or moving machines fabricated by human industry, and that with help of but few pieces compared with the great multitude of bones, muscles, nerves, arteries, veins, and other parts that are found in the body of each animal. Such persons will look upon this body as a machine made by the hands of God, which is incomparably better arranged, and adequate to movements more admirable than is any machine of human invention.”

For Descartes, there was no qualitative difference between a human made machine and an animal body: They are both entirely made up of matter extended in space, organized in terms of mechanism. Nature and technology had not yet parted ways, had not yet become different ontological realms. At the turn from the 17th to the 18th century this began to change, for two reasons. The first was the rise of empiricism: There was a nascent call, in the second half of the 17th century, especially among British philosophers, to provide an account of how humans can gain knowledge of the physical world that is not dependent –– as Descartes had it -- on the divine gift of reason but rather on experience. The second was the discovery, in the 1750s, of epigenesis and, subsequently, of the emergence of the concept of organisms: Up until that time, it had been assumed that embryos were fully formed miniature humans that would simply grow bigger over time. The recognition that this is not the case led to intense discussions about growth and the generation of matter, which in turn led to a philosophy of organisms and a sharp differentiation between nature and machines.

The perhaps most important author of this differentiation was Immanuel Kant. Physics or the study of matter, he argued, can easily be reduced to a set of mechanical causes as articulated by Newton, but [2] “Can we claim such advantages about the most insignificant plant or insect? Are we in a position to say: Give me matter and I will show you how a caterpillar will be created?”

“It is quite certain,” he explained in his 1790 Critique of Judgment, “that we cannot sufficiently get to know, still less explain, organized things and their inner possibility according to merely mechanical principles of explanation; so certain indeed that we can say boldly that it is absurd for human beings even merely to envisage such an attempt, or to hope that some day another Newton could arise who would make conceivable even so much as the production of a blade of grass according to natural laws which no intention has ordered”. The dividing line, for Kant, was growth or generation of form where before there was none [3]: “It would be absurd,” Kant declared, “to regard the initial generation of ... an animal as a mechanical effect incidentally arising from (matter).” Did this mean that animals could have intelligence and thus be free? No, it did not. In fact, the differentiation of organisms from machines, which had its origins in the final decades of the 17th century, was paralleled by a careful reworking of the difference between humans and animals: And it was in the context of this reworking that the term consciousness was invented: as the epitome of human-unique and human-exclusive freedom. 

The place where this invention of consciousness first happened was John Locke’s An Essay Concerning Human Understanding, which marked the birth of British empiricism. Locke (just like Kant more than half a century later) was in favor of Descartes’ conception of the human as the thinking and knowing thing. But he was uncomfortable with the fact that Descartes had suggested that what made it possible for humans to think and know was the divine gift of reason. To Locke, this was foolish: Grounding the capacity of human self-assertion against the unfounded authority of the clerics in a metaphysical argument meant to jeopardize the whole project launched by Descartes. What to do? How to provide a plausible possibility of knowledge without any recourse to metaphysical principles? Locke’s answer was his An Essay Concerning Human Understanding, published in 1689, at the center of which was a then radically new concept of mind. If mind up until then was a rarely systematically discussed religious concept – a kind of animating that referred to the Hebrew ruah and the Greek pneuma – then Locke re-purposed it as a depository of image-like ideas.

Plato and Aristotle – and libraries worth of authors after them – had argued that the way to true knowledge is contemplation: We (or at least a small set of free men) had access, through our souls, to the eternal ideas that organized reality.
Sensuous experience, in sharp contrast, was considered confusing at best. Here is Plato (Phaedo): “Now weren’t we also saying a while ago that whenever the soul additionally uses (proschrestai) the body to consider something (skopein ti), whether through (dia) seeing or hearing or some other sense (aisteseos) – for to consider something through the body is to consider it through a sense (aisteseos) – it is then dragged by the body to the things that are never the same; and it wanders about and is confused and dizzy, as if drunk, because the things it is grasping are of the same sort?”

Locke, following the changes brought about by Descartes but also by Thomas Hobbes and Robert Boyle, favored the exact opposite path: To him, knowledge consisted in the study of matter, in the investigation of empirical, material objects and their mechanical make-up. Consequently, the question concerning knowledge consisted in providing a plausible account of how an understanding of a given object, say a tree, makes it from the outside world into my understanding? His answer, famously, was sensuous experience: It is through a sensuous encounter with an object that small corpuscles transport the primary qualities of this object to your body, which then forms an idea of this object in our mind. This focus on the senses – which may sound banal today, but was a disturbingly novel idea at the time – was a sharp departure from Descartes, whose concept of knowledge was completely mental: Understanding of outside objects happened through pure reasoning in the soul.

The question Locke left largely unaddressed in the first edition of his Essay was: How does one have access to this gallery of images? In the second edition, published in 1694, he offered an at the time confusing new answer: consciousness (probably a translation of Descartes connaissance de soi or self-knowledge). Consciousness, said Locke, is the knowledge one has of oneself. It is not yet given at birth and hence is separate from the soul. Rather, consciousness is built up over time, as a cumulative self-awareness that defines one’s personal identity. The role of consciousness in Locke’s theory of knowledge was that the conscious self can examine – by way of thought – the ideas in one’s mind; it can perceive of true and false ideas, it can refine ideas through examination of objects, and it can discover connections between ideas and thereby build up: knowledge. 

And animals? In principle, said Locke, there is no difference between humans and animals: both are living, sensuous organisms that form experiences. It is just that humans, and only humans, have this one extra characteristic: consciousness.  Hence, humans are free – free to know themselves and the world – and animals not. No consciousness, no thought, no knowledge, no intelligence, no freedom. Or, in Kant’s words: humans have world and animals are part of the world.

It may sound confusing that the term consciousness was only introduced in the 1690s. Is consciousness not a perennial question? Well, historically it seems not. Before Locke consciousness was neither needed nor would it have made sense. Personal identity was granted by way of the soul and true knowledge was gained by contemplating first principles or ideas that were to be found in one’s soul. It was only once nature was reduced to matter and when recourse to the soul seemed to jeopardize the project of human self-liberation, that the question emerged as to how a human can gain knowledge of the outside world without referring to a soul – and only then did a concept like consciousness become necessary and plausible. Locke claimed that his project was an examination of “the internal operations of the mind.” However, it would be more appropriate to say that his examination invented the object he set out to study: the mind and, as a part of it, consciousness. What Locke looked for – and found – was a concept that would confirm the special status Descartes had assigned to the human, without referring to the soul.

The argument that machines have no senses, no mind, and no consciousness – that they lack the very basic necessities that would allow for freedom and knowledge – has been critical not only for articulating the concept of the human on which the modern world has been built; but also for the distinctive modern understanding of reality as composed of three ontological realities sui generis: the realm of the human (reason, freedom), the realm of nature (non-human, non-technical), and the realm of machines (the artificial, non-human, non-natural). Such a division of reality into three ontological independent fields (which to this day structures most universities) is completely unknown anywhere before the 18th century. But how come that the inventors of the modern concept of the human – the authors of the mechanization of the world – took it for granted that machines cannot think? Where did the idea that machines lack reason, mind, consciousness actually come from?

\subsection*{Why are machines thoughtless?}

In the early 17th century, at the time Descartes and Hobbes composed their works, the term “Mechaniks” was still knew and referred to an “inferior class.” Jonathan Sawday offers an excellent summary of the then prevailing understanding of machines, mechanics, and machine makers [4]: “Historically, ‘mechanics’ ... had long been understood as engaged in crafts that were ... demeaning in some (often unspecified) way. ... A ‘mechanical’ understanding of phenomena ... actually betokened a lack of mathematical or theoretical understanding.” In John Dee’s preface to Humphrey Billingsley’s 1570 translation of Euclid’s Geometry, Dee wrote that “A mechanicien, or a mechanical workman is he, whose skill is without knowledge of mathematical demonstration.” The background to this low estimation of things mechanical was largely due to Plato and Aristotle. Famously, Plato had argued in the Republic that technology – or techné, the practical art of making things – is devoid of knowledge, insofar as it consists simply in imitations of the visible rather than in an understanding of the first principles or ideas that give visible things their form. Artisans, as he saw it, worked off of confusing imitations of the real thing, the first principles. In his not exactly accessible language (Republic). “In the one section, the soul is forced to investigate from hypotheses, by using as images what were at a previous stage things imitated, not by working towards a first principle, but toward a conclusion. In the other section, by contrast, it moves from hypothesis toward a first principle which transcends hypothesis, but without the images of the earlier section, and so constructs its way of operating from the very forms by themselves.”

For Plato, artisans “make use of the visible forms ... without considering the actual things” and rely on “apprehension by images” – and thereby stay on the level of the confusions of the senses and mere resemblances. The reason for this was that, or so the Greek philosopher had it, they lacked that part of the soul that allowed to see and understand the eternal ideas (Republic). They were capable of dianoia, here best translated as practical know how, but not of noesis, the kind of thinking that participated in the divine nous (or mind) that organized reality. In short, for Plato, artisans, insofar as they lack reason, are not capable of seeing the truth and hence the things they build have no resemblance to truth.

Aristotle, referencing Plato’s philosophy of ideas, added that even where artisans build something not found in nature, they only build it as nature would build it if nature did build it (Physics, 199a9-199a19). “If a house, for example, had been a thing made by nature, it would be made in the same way as it is now by art; and if things made by nature were made not only by nature but also by art, they would come to be in the same way as by nature. The one, then, is for the sake of the other; and generally art completes what nature cannot bring to a finish, and in others imitates nature.”

Technology can only imitate nature or finish what nature left unfinished. The writings of Plato and Aristotle silently organized thinking about – and also the practice of – technology for over 2,000 years. They established that: (1) technical products have no reality of their own, independent from or set apart from nature. Technology, that is, was not yet considered an ontological realm – the realm of the artificial – in need of its own science. On the contrary, technology was understood to be dependent on nature and, ultimately, part of nature; (2) tool and machine makers lack in reason; (3) the knowledge necessary to produce technology had no truth potential: it was not useful to get to know the world or nature. So, the whole field of techné was understood to be without reason, knowledge or potential for truth.

Descartes’ simultaneously broke with this assessment – and held on to it. Or better: he simultaneously built on it and strategically revised it. He broke with it, insofar as he declared – against the then dominant scholastic tradition – the makers of machines to be fully endowed with reason. The background to this break with the scholastics was his own experience with Mechaniks. From 1607 to 1615, René Descartes attended the newly founded Jesuit college at La Fleche. He was trained in scholastic interpretations of Aristotle and was predestined to become one of the guardians of the Aristotelian nature cosmos of the late middle ages. However, Descartes appears not to have enjoyed the prospect of becoming a scholarly cleric: Shortly after college, the young Frenchman moved to the Netherlands and joined the protestant king in his battle against the Catholics. It appears that, upon arrival, Descartes was assigned to the engineering squad, where he was ordered to take classes with Isaac Beeckman, then one of the most well known machine makers of Europe. Beeckman introduced Descartes to the world of math and mechanism – and Descartes seems to have been as excited as he was confused. Beeckman, in contrast to Descartes, was considered a layman. He had never completed a scholastic curriculum and hence was not considered to be in a position to contemplate the first principles that organized the cosmos. And yet, Beeckman could explain natural phenomena better than any scholastic cleric Descartes had met.

Descartes – who later turned against Beeckman and insisted he had learned nothing from him –  was struck: the longer he worked with the engineer, the clearer it became to him that the old Aristotelian metaphysics he had studied in college could not be maintained. Clearly, the old idea that reason was unequally distributed – that artisans lacked reason – could not be maintained: and the idea that the world was organized by metaphysical first principles had to be replaced by a careful mathematical study of the mechanical principles that actually organized it. Hence, the famous opening of the Discourse on Method: “The power of judging well and of telling the true from the false—which is what we properly call ‘reason’ – is equal in all humans.” However, Descartes was not the first scholastic thinker who attended to techné. His vindication of machine makers had been prepared by Galileo Galilei (to whose work Beeckman introduced him in the 1620s, that is, at the time Descartes was working on the Discourse).

In the late 1590s, Galilei – deeply impressed by the work of Renaissance artisans like Brunelleschi, Alberto, and Leonardo – made it a habit to visit workshops to study the work of the machine makers. He was specifically interested in machines that could move weights, such as catapults, water wheels, and rotary winches, and began to work out a mathematical account of how these machines worked. Appropriately, he called his mathematical theory mechanism – and he instantly applied the insights he had gained from this new branch of mathematics to natural phenomena, especially the movement of the planets. Galileo’s invention of mechanism as a mathematical field – “On Mechanics” was published in 1600 – marked the beginning of a radical change in the perception of machines and those who make them. If before Galileo – and before Descartes’ encounter with Isaac Beckman – techné was considered to be without value for understanding how nature works, and machines therefore had no truth potential, then the two scholars changed this emphatically. The knowledge necessary for building machines was now increasingly considered the only plausible way for understanding – for knowing – nature: Mechanism became a practical as much as an abstract-mathematical project and emerged as the prime way of discovering truth in the scientific revolution.

The inversion is most fascinating: If for Plato and Aristotle and almost every philosopher who came after them, human-made things lacked truth potential because the artisans had no access to the abstract, metaphysical first principles that organized the world, then now human-made instruments became the first principles that had the power to explain the world: And building machines became the possibility of human self-assertion and knowledge.

Indeed, to explain the world mechanically meant to explain the world in terms of human-made machines and thus in quintessentially human terms (a significant break away from the old idea that the world was simply not accessible in human terms and that human-made things had no truth potential). The culmination point of explaining the world in human– in mechanical – terms were Newton’s laws (indeed, Newton’s mechanics was based, following Galileo’s lead, on a mathematical study of movement enabled by human-made instruments).

However, Descartes also held on to the old idea that techné or machines are without reason. It is almost as if he said: "But don’t you see, we all made a mistake. We have to revise everything. We thought that machine makers lacked reason because – as Plato and Aristotle put it – they built machines in terms of visible, physical reality. And it is true, that is what they do. It is just that we thought, following Plato and Aristotle, that this is the wrong thing: we blamed them and said that they lacked reason – the nous or the first principles that organize physical reality. But what if there are no first principles? Then Plato and Aristotle were simultaneously wrong and right. They were wrong, because there are no first principles – but they were also right because, indeed, physical reality is without reason; it is composed of automata, of matter organized in mechanical terms. And so if we want to understand it, we have to all become machine makers."

Descartes rescued the machine makers from Plato and Aristotle – but not the machines. In fact, that was his coup: By way of declaring that visible nature is all there is, that this visible nature is composed of machines, and that both nature and machines are without reason, Descartes was in a position to articulate the three critical interventions mentioned above:

\begin{enumerate}

\item He could break with the claims of the clerics that he world is organized in terms of metaphysical first principles: after all, mechanism could better explain the world than scholastic interpretations of Aristotelian texts.

\item He could argue that all humans are equally endowed with reason and can see the truth: after all, the knowledge of machinists was sufficient to explain nature (inevitably this means that what truth is and what reason is changed from a metaphysical into a physical conception).

\item He could insist that humans – endowed with reason – are obviously more than nature: after all, nature was organized in terms of mechanical and not of metaphysical principles or divine reason.

\end{enumerate}

\noindent Indeed, one could argue that Descartes’ revolutionary invention of a general, all inclusive concept of the human was contingent on making machine makers the exemplary humans – and that he could achieve this only by declaring nature to be composed of machines. For when Descartes writes that “reason is the only thing that sets us apart,” then what is the exemplary reference for the us? Arguably, the machine makers. Only humans, one could summarize Descartes, build machines and hence only humans are capable of knowledge. In short, Descartes’ concept of the human drew heavily on the old discourse on technology – but mutated it almost beyond recognition. From now on, machine and freedom – or intelligence and matter/mechanism – were considered mutually exclusive. If anything, the invention of the organism-machine distinction in the 18th century further cemented the conviction that the configuration that made possible freedom – that lifted humans and only humans above mere matter/mechanism – was not available to machines: experience + feeling + mind + consciousness + thought = intelligence/freedom.

Enter a 300 qubit based animat. We cannot help but ask, what if Descartes had said that reason – and hence consciousness and knowledge and freedom – were an intrinsic property of matter? What concept of the machine would that have made possible? And what concept of the human – of the role and place of the human in the world?

\subsection*{A philosophical laboratory} 

We have outlined an experiment for how to build machines in terms of the freedom quantum mechanics allows for. As we see it, classical Newtonian physics is very much indebted to mechanism as established by Galileo and Descartes. If one builds machines from this perspective, then freedom - consciousness and agency – are impossible.that  But what if one would build machines from the perspective of the freedom that quantum mechanics allows for: if one were to build machines in terms of the non-determinism of the world, in terms of panpsychism, and in terms of agency? 

The philosophical significance of the effort to build machines in terms of freedom is that it offers an opportunity to break not only with the concept of the machine that is at the center of the modern understanding of the human, but also with the modern comprehension of reality. The implications of this break are quite sweeping, because if our experiments were successful, that is, if an animat endowed with 300 qubits would indeed have consciousness and free will, that is, if it would have true agency, then this would require us to revise not only the concept of what a machine is but also the concept of the human and of nature that is built on the back of the modern machine concept. What is more, one would have to revise what we mean when we say animal, consciousness, freedom, intelligence, reality, technology, truth. Two obstacles are in the way of appreciating the philosophical stakes of our experiments.

\begin{enumerate}

\item  To argue that machines can be free and have consciousness may at first seem to suggest that machines can be human too: If only humans are free – then saying machines are free too means to grant machines a quality that was hitherto reserved for humans only.

\item
One may understand our suggestion that machines can be free and have consciousness as a reduction of humans to mere machines. For if consciousness were explainable in terms of machines then humans, counter to what we moderns thought thus far, could be sufficiently explained in terms of machines: Being human would be, in principle, no different from being a machine. And the language of engineering would be sufficient to account for the human.

\end{enumerate}

\noindent In our observation, both of these obstacles ultimately hold on to the concept of the machine and the human invented by Descartes and other early modern philosophers. To assume we are humanizing machines would inevitably mean making the concept of consciousness and freedom invented to consolidate human exceptionalism the model for all forms of consciousness and freedom. And to assume we would reduce humans to machines, would only make sense if we would hold on to the classical modern conception of the machine as without reason, without freedom.
It would mean holding on to the concept of matter and machines developed in terms of human-made instruments. Our work, however, radically calls these concepts of the machine and the human in question: And also the ontologies – the clear cut distinction between human things, natural things, and technical or artificial things – that were built on their back. If both freedom and consciousness – the freedom to consciously act – are intrinsic to matter then neither the argument that machines are necessarily devoid of consciousness would hold nor the argument that consciousness is a unique human quality that lifts us above mere matter, above mere mechanism. Three transformations, in particular, excite us.

\begin{enumerate}
\item 
{\bf The human}
The human would then no longer be the only, the defining instance of freedom and consciousness but one example among many: Human consciousness/freedom would be just one entry in a series; a series comprised of humans as well as of machines and animals and plants and microbes.That is, it would be an entry in a series that renders useless the modern ontology that strictly separates human things, natural things, and technical or artificial things.
\item 
{\bf The machine}
Machines would no longer be the exact opposite of reason, intelligence, freedom. They would no longer be instances of rogue, mindless, automated, repetitive behavior fully reducible to the intent of the engineer who built them.
No, machines built in terms of freedom would have an agency of their own, beyond human intent or control: and it is precisely this aspect that would make them valuable.
Machines could be something in themselves.
\item 
{\bf The practice of engineering.}
Ultimately, mechanism was an explanation of the world in terms of human-made instruments. It was a practice of simultaneously reducing the world to human intent and control. With our animat this inevitably changes: building machines in terms of quantum mechanics means to build machines in terms of the world (as brought into view by quantum mechanics) rather than of the human.
What is more, it means integrating machines into a fabric – into a reality – that is bigger than the human and predates the human by billions of years.
\end{enumerate}

\noindent These, then, are the philosophical stakes of building a 300 qubit based animat:
The experiments we propose will inevitably challenge and, likely, change what we mean – since the early modern period – when we say human, nature, technology, machine, freedom, consciousness, agency. The power of our qubit based animat is that it releases the world from the grip of the modern concept of the human/the machine and invites us to think again. That is, our work will likely change some of the very key terms that have defined what it is to be human for almost 400 years. It is no exaggeration to claim that the entire philosophical – conceptual – basis of modernity is at stake in the experiments we propose. What new vocabulary for being human will they give rise to? At present no-one knows. However, what we do know is that the old vocabularies have lost their validity. Experimentation, thus, is essential.

\subsection*{Coda: after the boring game}

It is a distinctive feature of contemporary technology – whether we talk about quantum computing or about microbiome research or synthetic biology – that it enables knowledge, often in the form of new practical possibilities, that render the classical modern understanding of the human/the world untenable: as untenable as the emergence of mechanical philosophy once rendered the Platonic and Aristotelian understanding of the human/the world. Call it a mutation of the space of possibility in which we have to organize our thoughts. A hugely fascinating consequence of this perhaps provocative but hard to argue with observation is that technology has become a philosophical laboratory. But where are the experimenters who can practice philosophy in terms of engineering? Who can simultaneously build a qubit based animat and understand how doing so may change the architecture of how we have been thinking?
Who can build machines in ways that reflect our new understanding of humans, of nature, of machines, of engineering? At present, they might not exist. That has much to do with the fact that our universities are, to this day, organized in terms of the old concept of the human and the machine. They have a faculty of arts (concerned with the human insofar as it is more than nature or a machine and hence cannot be explain in terms of the natural sciences or engineering); a faculty of science (concerned with nature as the non-human and as the opposite of the artificial); and a faculty of engineering (concerned with machines understood as the non-natural and the non-human). And to this day – call it the boring game – any attempt by the natural sciences or engineering to explain the human results in a defense of the human as more than nature/machine on the side of the faculty of arts.

Our animat makes this boring game implausible because it renders untenable the premises on which it is based. Is building machines in terms of freedom a philosophical or an artistic or a technical task? To understand what comes after the human/the machine we need experimenters – call it the work force of the future – who (1) understand that technology itself has become philosophical and (2) who can practice philosophy in terms of technical experiments. For our experiment, building on the work of the 'Transformations of the Human' program at the Berggruen Institute, we seek to address this need by composing a team made up of philosophers, artists, and engineers. Our hope is that our collaboration will give contours to a new, experimental and yet teachable expertise that unfolds beyond the old and now dysfunct division of labor between the human sciences and the natural sciences and engineering. An expertise that brings philosophy + art + technology together in a single practice, a practice that allows us to build machines that reflect an understanding of their philosophical stakes. Call it the work of freedom.

\subsubsection*{References for the Philosophical Commentary}

[1] René Déscartes. 1630. Discourse on Method. 
[2] Immanuel Kant. 1755. Universal Natural History and Theory of the Heavens.
[3] Immanuel Kant. 1763. The Only Possible Basis for a Demonstration of the Existence of God.
[4] Jonathan Sway. 2007. Engines of the Imagination. London: Routledge.

\end{document}